\documentclass[onefignum,onetabnum]{siamart190516}

\ifpdf
\hypersetup{
  pdftitle={Variance and covariance of distributions on graphs},
  pdfauthor={K. Devriendt, S. Martin-Guttierez and R. Lambiotte}
}
\fi

\usepackage{mathtools}
\usepackage{amsmath}
\usepackage{amssymb}
\usepackage{graphicx}
\usepackage{amsfonts}
\usepackage{xcolor} 
\usepackage{bbm} 
\usepackage{float} 
\usepackage{amsopn} 
\usepackage{mathrsfs} 
\usepackage{comment} 
\usepackage{tabularx} 

\renewcommand{\Pr}{\operatorname{Pr}}
\newcommand{\var}{\operatorname{var}}
\newcommand{\cov}{\operatorname{cov}}
\newcommand{\corr}{\operatorname{corr}}
\newcommand{\tr}{\operatorname{tr}}
\newcommand{\supp}{\operatorname{supp}}
\newcommand{\diag}{\operatorname{diag}}
\newcommand{\divg}{\operatorname{div}}
\newcommand{\argmax}{\operatorname{argmax}}
\usepackage{listings}
\definecolor{codegreen}{rgb}{0,0.6,0}
\definecolor{codegray}{rgb}{0.5,0.5,0.5}
\definecolor{codepurple}{rgb}{0.58,0,0.82}
\definecolor{backcolour}{rgb}{0.95,0.95,0.92}

\lstdefinestyle{mystyle}{
    backgroundcolor=\color{backcolour},   
    commentstyle=\color{codegreen},
    keywordstyle=\color{magenta},
    numberstyle=\tiny\color{codegray},
    stringstyle=\color{codepurple},
    basicstyle=\ttfamily\footnotesize,
    breakatwhitespace=false,         
    breaklines=true,                 
    captionpos=b,                    
    keepspaces=true,                 
    numbers=left,                    
    numbersep=5pt,                  
    showspaces=false,                
    showstringspaces=false,
    showtabs=false,                  
    tabsize=2
}

\usepackage{lipsum}
\usepackage{amsfonts}
\usepackage{graphicx}
\usepackage{epstopdf}
\usepackage{algorithmic}
\ifpdf
  \DeclareGraphicsExtensions{.eps,.pdf,.png,.jpg}
\else
  \DeclareGraphicsExtensions{.eps}
\fi

\newsiamremark{remark}{Remark}
\newsiamremark{hypothesis}{Hypothesis}
\crefname{hypothesis}{Hypothesis}{Hypotheses}
\newsiamthm{claim}{Claim}

\headers{Variance and covariance of distributions on graphs}{K. Devriendt, S. Martin-Guttierez, and R. Lambiotte}

\title{Variance and covariance of distributions on graphs\thanks{Accepted for publication in SIAM Review, Research Spotlights
\funding{
KD was supported by The Alan Turing Institute under the EPSRC grant EP/N510129/1. SMG acknowledges support by the Spanish Ministry of Science, Innovation and Universities (MICIU) under Contract No. PGC2018-093854-B-I00 and Spanish Ministry of Education, Culture and Sport (Grant No. FPU15/01461).}}}

\author{Karel Devriendt\thanks{Mathematical Institute, University of Oxford, Oxford, UK and The Alan Turing Institute, London, UK
  (\email{karel.devriendt@maths.ox.ac.uk})}
\and Samuel Martin-Gutierrez\thanks{Grupo de Sistemas Complejos, Universidad Polit\'{e}cnico de Madrid, Madrid, Spain and Complexity Science Hub Vienna, Wien, Austria
  }
\and Renaud Lambiotte\thanks{Mathematical Institute, University of Oxford, Oxford, UK}}

\begin{document}

\maketitle

\begin{abstract}
We develop a theory to measure the variance and covariance of probability distributions defined on the nodes of a graph, which takes into account the distance between nodes. Our approach generalizes the usual (co)variance to the setting of weighted graphs and retains many of its intuitive and desired properties. Interestingly, we find that a number of famous concepts in graph theory and network science can be reinterpreted in this setting as variances and covariances of particular distributions. As a particular application, we define the \emph{maximum variance problem} on graphs with respect to the effective resistance distance, and characterize the solutions to this problem both numerically and theoretically. We show how the maximum variance distribution is concentrated on the boundary of the graph, and illustrate this in the case of random geometric graphs. Our theoretical results are supported by a number of experiments on a network of mathematical concepts, where we use the variance and covariance as analytical tools to study the (co-)occurrence of concepts in scientific papers with respect to the (network) relations between these concepts.
\end{abstract}




\section{Introduction}\label{S: Introduction}
The variance of a probability distribution is a fundamental concept in the toolkit of probability theory and statistics and is routinely applied throughout science, engineering and numerous practical settings. Intuitively speaking, the variance captures how spread-out the outcomes of a distribution are, and thus reflects the inherent variability in this distribution. In many practical cases however, probability distributions are defined on the nodes of a network: websites on the internet, individuals in a social network, neurons in the brain, etc. These nodes are the building blocks of a network, and when studying distributions or signals defined on nodes it is natural to take the underlying network structure into account. As the usual definition of variance can not take this structure into account, we thus lack a basic methodological tool when analysing distributions and signals on a graph.

In this article, we propose a measure of variance and covariance for distributions defined on a network, which take into account the underlying structure of the network by considering the distances between nodes. These distances provide a notion of what it means to be `spread out' on the network, which in turn allows us to define (co)variances of distributions on the network. Our proposed formulas for variance and covariance take a very simple mathematical form (as a quadratic product and matrix trace, respectively) yet still capture many of the intuitive and mathematical properties of the usual (co)variance. To illustrate our new measures in practice, we apply the proposed variance and covariance measures to the analysis of an empirical network of mathematical concepts with data from a collection of scientific papers. Our approach allows for a unified and intuitive treatment of the structural (relations between concepts) and functional data (usage of concepts in papers) in this system and we describe some qualitative and quantitative findings.
As a second application, we show that the variance and covariance of some particular distributions correspond to previously known graph characteristics, offering a new framework to interpret and understand them.

We furthermore consider the \emph{maximum variance problem} which seeks to determine the largest possible variance on a given graph, and to characterize the distribution(s) that attains it. As a theoretical contribution, we find a complete characterization of the maximum variance distribution when considering the \emph{effective resistances} as a distance measure between nodes. We show that this maximum variance distribution is concentrated on the \emph{boundary} nodes of a graph, and provide support for this intuition by experiments on random geometric graphs and further analytical results on a number of simple graphs.

To the best of our knowledge, the measures that we introduce for variance and covariance are new in the context of distributions on networks. Mathematically, our variance measure is a special case of the well-known Rao's quadratic entropy \cite{Rao, Stirling} which measures the diversity of distributions on general categorical variables with a notion of dissimilarity between the variables. This correspondence connects our work to the more general setting of \emph{diversity measures} which are widely used, for instance in ecology \cite{Schleuter_functional_diversity, Cadotte_diversity}. The work of Leinster \emph{et al.} on maximum diversity distributions \cite{Cobbold_Leinster, Leinster_book} is particularly relevant (see Section \ref{S: maxvar distributions main}). Additionally, the effective resistance matrix is well-studied in the context of graph theory \cite{Bapat_book, Klein_resistance_distance, Zhou_on_resistance_matrix_node-transitive} and our variance quadratic form \eqref{eq: definition resistance variance and covariance} and the related maximum variance problem \eqref{eq: maximum variance problem} were defined and studied before in different contexts. Hjorth et al. \cite{Hjorth_finite_metric_spaces} studied expression \eqref{eq: maximum variance problem} as a generalized diameter of metric spaces, and described the conditions for which this problem has a unique solution. Dankelmann \cite{Dankelman_average_distance} defines the quadratic form in \eqref{eq: definition resistance variance and covariance} for the geodesic graph distance and solves the corresponding maximization problem for tree and cycle graphs. These results are generalized from the geodesic graph distance to the effective resistance in \cite{Bapat_Neogy_Maxvar_on_trees, Dubuy_Neogy_Solving_quadratic_program}, where it is shown that the corresponding maximum variance problem can be solved efficiently.

The rest of this article is organized as follows: In Section \ref{S: Variance and covariances on graphs} we introduce the relevant mathematical background about graphs and (joint) distributions on graphs and introduce our new variance and covariance measures. Section \ref{S: (co)variance in a network of knowledge} treats an application of the (co)variance measures to a network of mathematical concepts. In Section \ref{S: correspondences between (co)var and graph measures}, we show how our new measures relate to a number of existing concepts. In Section \ref{S: maxvar distributions main} we introduce and solve the maximum variance problem on a graph, with respect to the effective resistance distance. We characterize the maximum variance solutions, with a particular focus on the support of the maximum variance distributions. Section \ref{S: conclusion} concludes the article with a summary of the results and an outlook to further possible applications.
\section{Variance and covariance on graphs}\label{S: Variance and covariances on graphs}
\subsection{Preliminaries}\label{SS: preliminaries}
We start by introducing a number of preliminary notions about graphs and probability theory. A \emph{graph} $G$ consists of a set of $n$ nodes $\mathcal{N}$ and a set of links $\mathcal{L}$ that connect pairs of nodes; a link between two nodes $i$ and $j$ thus corresponds to an element $(i,j)\in\mathcal{L}$. Every link is furthermore assigned a (positive, real) weight $c_{ij}>0$, resulting in a \emph{weighted graph}. The \emph{degree}\footnote{The number of neighbours of a node $i$ will be called the combinatorial degree to distinguish it from the `weighted' degree $k_i$} of a node is defined as the combined weight of all links connected to a node, as $k_i=\sum_{j}c_{ij}$. Since graphs can be seen as weighted graphs where $c_{ij}=1$ for all links, we will further work in the more general setting of weighted graphs. Furthermore, we will assume all graphs to be finite $(n<\infty)$ and connected, which means that there is a path between every pair of nodes. 
\\
A \emph{distribution on a graph} is a function $p:\mathcal{N}\rightarrow [0,1]$ which assigns a nonnegative number $p(i)$ to each node in the graph, such that all numbers add up to one $\sum_{i\in\mathcal{N}}p(i)=1$. Distributions can be used to define a \emph{random node} $N$, which is a random element of the node set, with probability to equal any of the nodes $i$ given by the corresponding distribution $p(i)$. In other words, we can `sample' the random node $N$ and get any of the nodes of $G$ with probability
$$
\Pr[N=i] = p(i) \text{~for all~}i\in\mathcal{N}.
$$
For this reason, the value $p(i)$ of an element is called the \emph{probability} of $i$. To specify the underlying distribution of a random node $N$, we often write $N\sim p$ and say that $N$ is distributed according to $p$.
\\
A \emph{joint distribution on a graph} is a function $P:\mathcal{N}\times\mathcal{N}\rightarrow [0,1]$ which assigns a nonnegative number $P(i,j)$ to each pair of nodes in the graph, such that all numbers add up to one. A joint distribution can be used to define a random pair of nodes $(N,M)$ which are two random elements of the node set, with probability of being sampled given by the joint distribution as
$$
\Pr[(N,M)=(i,j)]=P(i,j)\text{~for all~}(i,j)\in\mathcal{N}\times\mathcal{N}.
$$
Any joint distribution $P$ on node pairs also naturally leads to two (simple) distributions $\tilde{p}$ and $\tilde{q}$ on the nodes, defined by $\tilde{p}(i)=\sum_{j\in\mathcal{N}}P(i,j)$ and $\tilde{q}(j)=\sum_{i\in\mathcal{N}}P(i,j)$, which are called the \emph{marginal distributions} of $P$.  
\\
In Section \ref{S: (co)variance in a network of knowledge} we consider an application of these concepts in a practical setting, which serves as an illustration of the definitions above. We will consider a graph made up of nodes that represent mathematical concepts with links that reflect conceptual relations between pairs of concepts (inferred from Wikipedia hyperlinks) and study a collection of scientific papers that use these concepts. The occurrence of a subset of concepts in a given paper is translated to a distribution on the relevant nodes (concepts) in the network, and the frequency of pairs of concepts appearing together in a paper is represented as a joint distribution on the network. 
\subsection{(co)variance with respect to node distances}\label{SS: defining the (co)variances}
\begin{figure}[h!]
    \centering
    \includegraphics[scale=0.2]{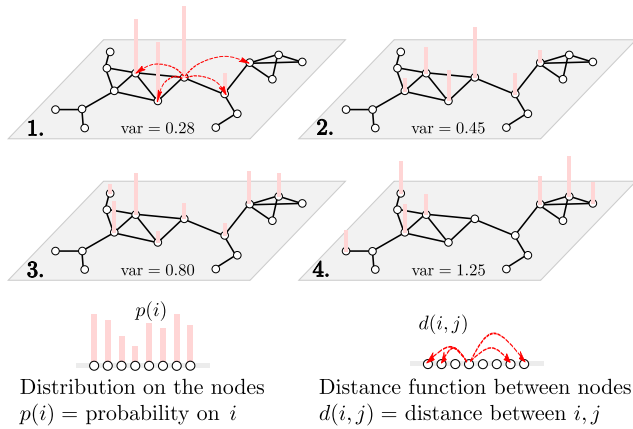}
    \caption{The variance of distributions on a graph can be defined with respect to a distance function between the nodes of the graph. This measure allows us to compare how `spread out' different distributions are on the network. In the example above, the distributions become more spread out over the network, going from distribution $1.\rightarrow 4.$ with an increasing variance as a result; here, the variance is calculated with respect to the square root resistance distance as in equation \eqref{eq: definition resistance variance and covariance}.
    }
    \label{fig: illustration of distance and variance on network}
\end{figure}

As introduced, the interpretation of a graph distribution is twofold: we can consider it as a signal (function) on the nodes of a graph (see also \cite{Shuman_graph_signal_processing}), or as representing a random node. In both cases, it is natural to ask whether a distribution is centered, or concentrated, on a small part of the graph and thus might be well understood and described by restricting our attention to this small part, or if instead the distribution is spread out over the graph. In the setting of distributions on the real numbers (or other vector spaces), the variance is a natural quantifier of exactly these properties; it reflects how spread out a distribution is as the average squared difference between a random outcome of the distribution and the mean. To generalize this standard notion of variance to graphs, we propose to take into account distances between nodes in the graph, i.e. a function $d:\mathcal{N}\times\mathcal{N}\rightarrow \mathbb{R}$ that says something about `how far' $d(i,j)$ two nodes $i$ and $j$ are in the graph. Given such a distance notion between nodes, we propose a generalization of variance to distributions on a graph:
\begin{equation}\label{eq: definition graph variance}
\var(p) = \frac{1}{2}\sum_{i,j\in\mathcal{N}}p(i)p(j)d^2(i,j)
\end{equation}
As shown in Appendix \ref{A: standard (co)variance as special case}, definition \eqref{eq: definition graph variance} is a proper generalization of the usual variance, which is retrieved if $\mathcal{N}\subset\mathbb{R}$ with the usual Euclidean distance. We will later discuss a number of examples of distance functions that illustrate how expression \eqref{eq: definition graph variance} might be used in practice. Figure \ref{fig: illustration of distance and variance on network} illustrates how our notion of variance captures the variance of a graph distribution.
In the case of joint distributions (i.e. pairs of random variables), the covariance is used much in the same way as the variance to quantify whether random pairs sampled according to this distribution are on average close together, or far apart. Making use of a distance notion $d$ again, we propose a generalization of the covariance to joint distributions $P$ on a graph:
\begin{equation}\label{eq: definition graph covariance}
\cov(P) = \frac{1}{2}\sum_{i,j\in\mathcal{N}}\left[\tilde{p}(i)\tilde{q}(j)-P(i,j)\right]d^2(i,j)
\end{equation}
Appendix \ref{A: standard (co)variance as special case} we show that equation \eqref{eq: definition graph covariance} indeed generalizes the usual covariance, which is retrieved if $\mathcal{N}\subset\mathbb{R}$ with the usual Euclidean distance. 
\\
In the special case where $(N,M)$ are independent random nodes with distributions $p$ and $q$ respectively, the joint distribution equals $P(i,j)=p(i)q(j)$. As a consequence we retrieve $\cov(P)=0$ for independent random nodes. In the case where one random variable is a copy of the other, the joint distribution is a diagonal matrix with distribution the distribution of the random node on the diagonal, and we find $\cov(P)=\var(p)$. For some more examples of joint distributions and their covariance, see Appendix \ref{A: example of covariance}.
\subsection{Distance functions on graphs}\label{SS: distance functions on graphs}
Definitions \eqref{eq: definition graph variance} and \eqref{eq: definition graph covariance} measure the variance and covariance of distributions on a graph with respect to a certain `distance' between the nodes. The most famous distance on graphs is the shortest-path distance (or geodesic distance) where $d(i,j)$ is the length\footnote{In the case of unweighted graphs, the length of a path equals the number of links contained in the path. In the case of weighted graphs, the length of a path is defined as the sum $\sum\limits_{(i,j)\in\mathcal{P}}c^{-1}_{ij}$ over all links $\mathcal{P}$ in the path. From this perspective, the weight between two nodes acts as an affinity, where larger weights correspond to `better connected' nodes.} of the shortest path between two nodes $i$ and $j$. In addition to capturing the intuitive notion of a distance between nodes, the geodesic distance also satisfies the mathematical properties of a metric \cite{Harary_distance_in_graphs}.
\\
Another important metric between the nodes of a graph is the \emph{effective resistance} \cite{Klein_resistance_distance, Bapat_book, Ellens_effective_graph_resistance, Qiu_Clustering_commute_times}. Similar to the geodesic distance, the effective resistance reflects the length of the paths between a pair of nodes. However, instead of only taking the shortest path into account, the effective resistance is influenced by all paths (and their lengths) between a pair of nodes, and becomes smaller as more paths are available. Due to this more integrative notion of distance and its nice mathematical properties, the effective resistance is often preferred over the shortest-path distance when studying networks. We write $\omega_{ij}$ for the effective resistance between two nodes $i$ and $j$, and define this resistance based on the Laplacian matrix of a graph. The Laplacian matrix $Q$ of a graph with $n$ nodes is an $n\times n$ matrix with entries $(Q)_{ii}=k_i$ on the diagonal, $(Q)_{ij}=-c_{ij}$ for all links $(i,j)$ and zero otherwise, and can be used to define the effective resistance as
$$
\omega_{ij} = (e_i-e_j)^TQ^\dagger(e_i-e_j),
$$
where the unit vectors have entries $(e_i)_k=1$ if $k=i$ and zero otherwise, and where $Q^\dagger$ is the Moore-Penrose pseudoinverse of the Laplacian. For our application, the most relevant property is that both $\omega$ and its square root $\sqrt{\omega}$ are metrics between the nodes of a network \cite{Klein_resistance_distance, karel_resistance>distance}. With the choice of the square root effective resistance as a distance between nodes, the variance and covariance can be written in matrix form as
\begin{equation}\label{eq: definition resistance variance and covariance}
\var_{\omega}(p) = \tfrac{1}{2}\mathbf{p}^T\Omega \mathbf{p}\qquad\cov_{\omega}(P)=\tfrac{1}{2}\left[\tilde{\mathbf{p}}^T\Omega\tilde{\mathbf{q}} - \tr(P\Omega)\right]
\end{equation}
with the $n\times n$ matrix $\Omega$ containing the effective resistances as entries $(\Omega)_{ij}=\omega_{ij}$, and with $\mathbf{p}=(p(i),\dots,p(j))^T$ the vector containing the probability for all nodes, and similarly for $\tilde{\mathbf{p}},\tilde{\mathbf{q}}$ and matrix $(P)_{ij}=P(i,j)$ containing the probabilities of all pairs of nodes. We will further also use $p_i$ to denote an entry of the probability vector $\mathbf{p}$, which thus equals the probability $p(i)$. Replacing $\Omega$ by another distance matrix $(D)_{ij}=d^2(i,j)$ gives the general matrix form for variance and covariance.
\section{Variance and covariance in a network of knowledge}\label{S: (co)variance in a network of knowledge}
As an example application, we study a `network of knowledge' made up of mathematical ideas and results with links between related concepts. The code and data of our analysis are available on GitHub \cite{github}. We consider a list of mathematical concepts (theorems, lemmas, equations) compiled from four Wikipedia pages that list these concepts, and we infer links between the concepts from hyperlinks between their respective Wikipedia pages. More information about the data retrieval and filtering of the data set can be found in \cite{Salnikov_Renaud_network_of_knowledge}, where (a higher-order variant of) this network was investigated. The resulting network of concepts consists of $n=1150$ nodes and $m=4109$ links and is shown in Figure \ref{fig: network of knowledge} below.
\begin{figure}[h!]
\begin{centering}
\includegraphics[scale=0.28,trim=0cm 1cm 0cm 1cm ,clip]{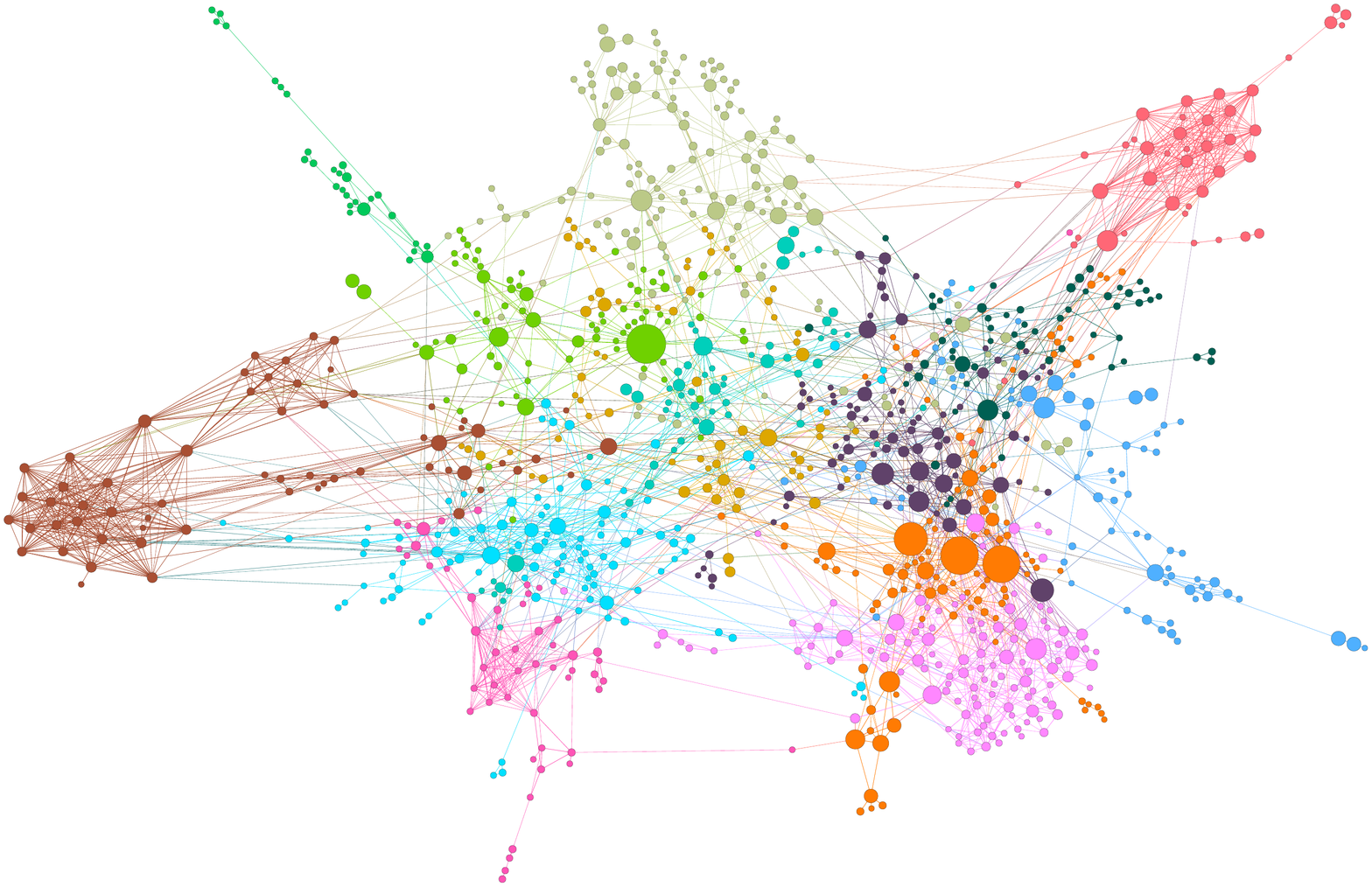}
\caption{Hyperlink network of Wikipedia pages of the considered mathematical concepts (see \cite{Salnikov_Renaud_network_of_knowledge}). The size of the nodes is proportional to their PageRank and the color coding corresponds to communities found using the Louvain algorithm.}
    \label{fig: network of knowledge}
\end{centering}
\end{figure}
We consider this network as the underlying structure of mathematical concepts and use it to investigate how these concepts are used in practice by their occurrences in scientific papers.
\\
To study the functional aspect of the network of knowledge, we use a corpus of $140$k+ papers from the arXiv and the mathematical concepts used therein. For each paper we count which of the mathematical concepts appear and represent this by a uniform distribution over the used concepts. Every paper $i$ thus has a corresponding subset of concepts $\mathcal{V}_i$ and distribution $p^{(i)}$ uniform over this set of concepts.

A first question we consider is whether the mathematical papers contain `coherent' sets of mathematical concepts. In terms of variance, this question can be addressed by comparing the variance of the paper distributions $p^{(i)}$ with a null model, representing `virtual papers'. Figure \ref{fig:empirical_variances_compared_to_null_model__and__subfields_variance} below shows that the paper distributions generally have a smaller variance, compared to virtual papers made up of randomly sampled concepts according to their relative frequency over the full corpus. This observation is confirmed by performing the one-sided Mann-Whitney U test, from which we find with high significance (p-values $<1\text{e-}12$) that the variance of a paper is typically smaller than what would be expected from the null model; more details on the test results are given in the Appendix \ref{A: statistics of experiment}. 
Intuitively, this observation reflects the idea that the practical use of mathematical concepts is related to the underlying connections between concepts, where a group of concepts is more likely to be considered in a paper if this group is coherent, as measured by a small variance of their corresponding uniform distribution. Furthermore, Figure \ref{fig:empirical_variances_compared_to_null_model__and__subfields_variance} clearly indicates a range of `typical' variance values which could be used to identify papers with exceptionally small (or high) variances. As a second application, the variance of paper distributions $p^{(i)}$ could be used as grounds for (qualitative) comparison between different fields of study. In the bottom plots of Figure \ref{fig:empirical_variances_compared_to_null_model__and__subfields_variance} for instance, we rank the different sub-fields based on their variance distributions, where the distributions on the top left are concentrated on smaller variances than the distributions on the bottom right, as quantified by one-sided Mann-Whitney U tests (see GitHub \cite{github} for the test statistics and p-values). More generally, this type of comparative analysis might be useful when there are different `modes of operation' of a single network, giving rise to different functional signals.

\begin{figure}[htbp]
\includegraphics[width=.99\textwidth]{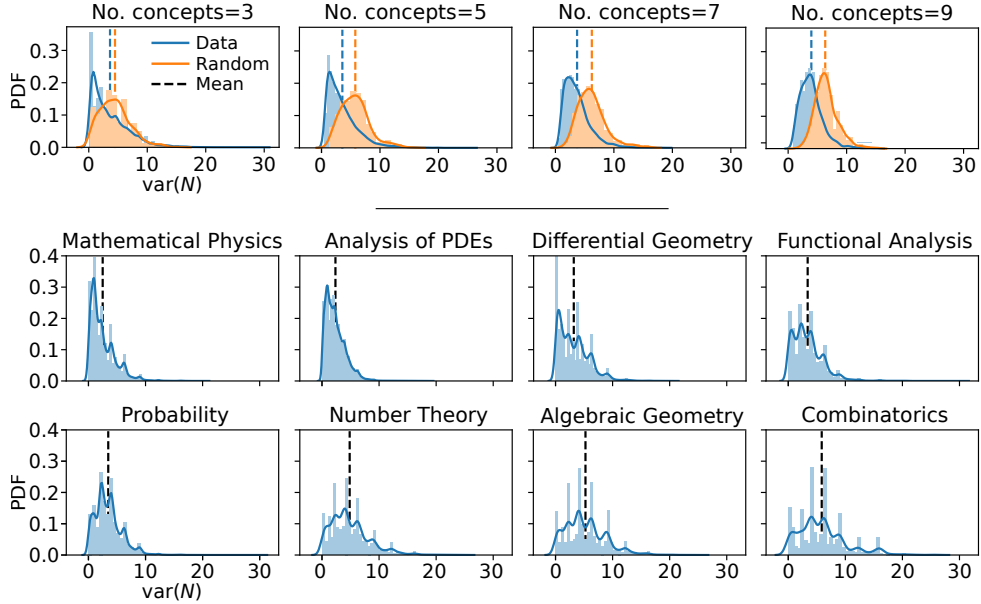}
\caption{(Top figure) Distribution of network variances of concept distributions $p^{(i)}$ in papers, calculated with respect to the geodesic distance in the Wikipedia network. In each panel, the probability density function (pdf) of variances is calculated for papers containing 3 (resp. 5,7,9) concepts present in the Wikipedia network, and is compared with the variance pdf for a collection of (null model) virtual papers with the same number of concepts. The empirical variance distributions are concentrated on smaller variances compared to the null model variances, as confirmed by the one-sided Mann-Whitney U test (see Appendix \ref{A: statistics of experiment}). (Bottom figure) Distribution of network variances of concept distributions $p^{(i)}$ for arXiv papers from different sub-fields. The fields are ordered from top left to bottom right according to concentration on increasing variances as found from the Mann-Whitney U test.}
\label{fig:empirical_variances_compared_to_null_model__and__subfields_variance}
\end{figure}
\begin{figure}
    \begin{centering}
    \includegraphics[width=0.99\textwidth]{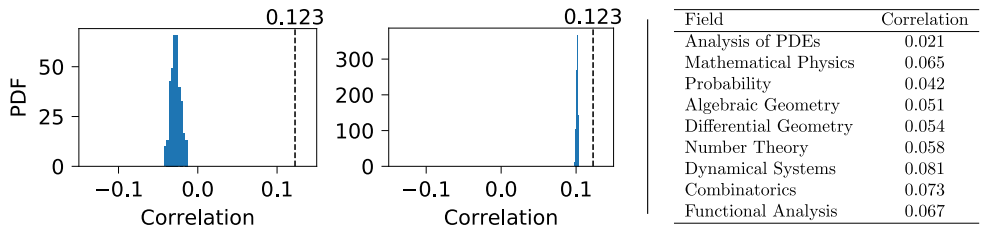}
    \caption{(Left figure) Comparison between empirical covariance of the co-occurrence distribution $P$ on the Wikipedia network of concepts and two null models. In the left panel, the covariance of $P$ is calculated with respect to the shortest-path distance on the Wikipedia network (dashed line) and the shortest-path distance on $100$ realizations of a degree-preserving randomization of the network (blue bars). In the right panel, the covariance of $P$ (dashed line) and of $100$ realizations of a marginal-preserving randomization of $P$ (blue bars) are calculated with respect to the shortest-path distance on the Wikipedia network. (Right table) Correlation of the joint probability distribution of concepts for papers of different sub-fields.}
    \label{fig:covariance_wiki_rnd__and__tab:covariances_fields}
    \end{centering}
\end{figure}

On an aggregate level, we can study the corpus of papers by counting the co-occurrences of pairs of concepts over all papers. This gives a joint distribution $P$, where $P(i,j)$ is proportional to the frequency of co-occurrence of concepts $i$ and $j$. To test how the function of the network of knowledge relates to its structure, we compare the covariance of this empirical distribution $P$ on the given graph against two null models: in the first model the distribution $P$ is left constant but the underlying graph is randomized, while in the second model the distribution $P$ is randomized while the graph is kept constant. Since we are comparing joint distributions with potentially different marginals, we consider a normalization of the covariance, which corresponds to the well-known concept of \emph{correlation}:
\begin{equation*}
    \corr(P) = \frac{\cov(P)}{\sqrt{\var(\tilde{p})\var(\tilde{q})}}
\end{equation*}
Since the co-occurrence joint probability distribution is symmetric, we have $\tilde{p}=\tilde{q}$, which means that the correlation is simply calculated as $\cov(P)/\var(\tilde{p})$.
\\
As a first null model, we perform a degree-preserving rewiring \cite{Mihail_rewiring} of the Wikipedia network while keeping the joint distribution constant. As seen in the left panel of Figure \ref{fig:covariance_wiki_rnd__and__tab:covariances_fields}, measuring the covariance of $P$ with respect to these randomized graphs yields significantly lower correlation values; this would imply that the empirical correlation is not simply a consequence of the (local) degree distribution of the network. Our second null model consists of a marginal-preserving randomization of $P$ while leaving the network intact: we pick two pairs of nodes $(i,j)$ and $(i',j')$ with non-zero joint probabilities and reshuffle their joint probabilities as
$$
P\rightarrow P-\alpha(v_{ij}v_{ij}^T+v_{i'j'}v_{i'j'}^T-v_{ij'}v_{ij'}^T-v_{i'j}v_{i'j}^T)
$$
where $v_{ij}=(e_i+e_j)(e_i+e_j)^T$ and with a uniform random value $\alpha<\min(P(i,j),\allowbreak P(i',j'))$. In other words, there is a shift of probability mass $\alpha$ from $(i,j)\rightarrow(i,j')$ and from $(i',j')\rightarrow(i',j)$. We repeat this procedure until all pairs of nodes have been involved, which produces a (symmetrically) randomized joint distribution $P'$ with the same marginals as $P$. The right panel in Figure \ref{fig:covariance_wiki_rnd__and__tab:covariances_fields} shows that these randomized joint distributions are concentrated on significantly lower correlation values. We remark that the empirical covariance is highly atypical for both null model classes and it is thus likely that these models are discarding too much structure to give a reliable baseline for the empirical covariance. A further development of appropriate null models for measuring the covariance on a network would be an interesting line of further research. We also calculated the co-occurrences for a number of sub-fields separately and report their correlations in the table on the right in Figure \ref{fig:covariance_wiki_rnd__and__tab:covariances_fields}. 

The above analysis illustrates a practical scenario where data is available in different modalities (i.e. structural and functional) and for which our variance and covariance measures enable a unified treatment of this system data. Our framework is of course not restricted to this specific example, but can be applied in the context of many other network problems in which a combination of structural and functional modalities are important.
\section{Correspondence between (co)variance and existing graph measures}\label{S: correspondences between (co)var and graph measures}
Apart from using the variance to quantify the spread of distributions in real-world applications, we can also further study equations \eqref{eq: definition graph variance} and \eqref{eq: definition graph covariance} theoretically. A first interesting result is that our proposed variance and covariance measures correspond to a number of existing concepts in graph theory and network science. This provides a new, probabilistic perspective on these concepts which adds to their understanding.
\subsection{Kemeny's constant}\label{SS: Kemeny}
A \emph{random walk} on a graph is a process in which a `random walker' makes its way across the nodes of a network by following the links in the network with certain probabilities \cite{Peres_Levin_Markov_Chains, Lovasz_randomwalks, Doyle}. More precisely, a random walk describes a sequence of random nodes $\lbrace N_t\rbrace_{t\in\mathbb{N}}$ where the consecutive random nodes $N_t$ - which represent the position of the random walker at timestep $t$ - are related via a transition matrix $T$ as
$$
\Pr(N_{t+1}=j \vert N_{t}=i) = T_{ji} = c_{ij}/k_{i}
$$
In other words, the probability that the walker leaves a node via a particular link is proportional to the weight of that link\footnote{Other types of random walks are possible and are defined based on different transition probabilities.}. For this type of random walker on a connected graph, the distribution of $N_t$ converges to a unique stationary distribution $\pi$, with probabilities $\pi(i)=k_i/(2m)$ where $m=\tfrac{1}{2}\sum_{i=1}^n k_i$, independently of the distribution of the initial position of the random walker $N_0$.
\\
In \cite{Kemeny_1981}, Kemeny discovered that the time $\kappa_i$ it takes on average for a random walker to go from node $i$ to a random node $J\sim \pi$ (with the stationary distribution) is independent of the node under consideration; in other words $\kappa_i=\kappa$ is a constant independent of $i$, which is now called Kemeny's constant. Following a recent result on Kemeny's constant in \cite{Wang_Kemeny_constant}, we find the following relation to a graph variance:
\begin{proposition}\label{proposition: Kemeny's constant}
Kemeny's constant $\kappa$ of a random walker is proportional to the steady-state variance of that random walker, measured with respect to the square root effective resistance, as
$$
\kappa = 2m\var_{\omega}(\pi)
$$
\end{proposition}
\textbf{Proof:} Proposition \ref{proposition: Kemeny's constant} follows from expression \eqref{eq: definition resistance variance and covariance} for the graph variance with respect to the square root effective resistance, and the relation between Kemeny's constant and the resistance matrix 
$\kappa = m\pi^T\Omega\pi$, as derived in \cite{Wang_Kemeny_constant}.\hfill$\square$
\\
One immediate consequence of Proposition \ref{proposition: Kemeny's constant} and the interpretation of Kemeny's constant as a variance is that it implies an upper-bound of $\kappa\leq 2m \sigma^2$, using the maximum variance results from Section \ref{S: maxvar distributions main}.
\subsection{Kirchhoff index/effective graph resistance}\label{SS: Kirchhoff index}
Apart from serving as a graph metric, the effective resistance has been used in a variety of applications in different domains. An important example is the sum of all effective resistances
$$
R_G=\frac{1}{2}\sum_{i,j\in\mathcal{N}}\omega_{ij}
$$
which is used in mathematical chemistry as a `fingerprint' of a network that represents a chemical compound - here, $R_G$ is called the Kirchhoff index \cite{Klein_resistance_distance, Palacios_Kirchhoff_index} - and is used as a heuristic measure of robustness and connectedness in the characterization of (empirical) networks - here, it is called the effective graph resistance \cite{Ellens_effective_graph_resistance, Ghosh}. Again, from expression \eqref{eq: definition resistance variance and covariance} for the graph variance, we immediately find the following correspondence:
\begin{proposition}\label{proposition: Kirchhoff index}
The Kirchhoff index or effective graph resistance $R_G$ of a graph is proportional to the variance of the uniform distribution on that graph, measured with respect to the square root effective resistance, as
$$
R_G = n^2\var_\omega(u/n)
$$
\end{proposition}
\textbf{Proof:} Inserting the uniform distribution $(u/n)(i)=1/n$ for all $i$, in equation \eqref{eq: definition graph variance} equals the formula for $R_G$ up to the constant $n^2$, which proves Proposition \ref{proposition: Kirchhoff index}.\hfill$\square$
\\
We remark that the uniform distribution also appears naturally as the stationary state of diffusion processes on a graph, see Appendix \ref{A: variance for diffusion}.
\subsection{Network modularity}\label{SS: modularity and Markov stability} Similar to the variance, we find that formula \eqref{eq: definition graph covariance} for the graph covariance is related to some known expressions in graph theory and its applications.
\\
In the network science community there has been a large and sustained research effort in developing methods to identify groups of nodes which are tightly connected within but poorly connected between, i.e. to uncover the so-called \emph{community structure}. Many results on this problem have been centered around the concept of \emph{network modularity} \cite{Newman_Networks}, which assigns a `score' $M(g)$ to each partition $g:\mathcal{N}\rightarrow\lbrace 1,\dots,k\rbrace$ of the nodes into $k$ communities, as
$$
M(g) = \frac{1}{2m}\sum_{i,j\in\mathcal{N}}\left((A)_{ij} - \frac{k_ik_j}{2m}\right)\delta_{g(i)g(j)},
$$
with adjacency matrix $A$ which has entries $(A)_{ij}=c_{ij}$ for all links, and with the Kronecker delta defined as $\delta_{g(i)g(j)}=1$ if and only if $i$ and $j$ are in the same group. In Appendix \ref{A: modularity and markov stability} we show that the modularity function $M$ can be interpreted as the covariance of an appropriate joint distribution and distance function:
\begin{proposition}\label{propos: correspondence with modularity}
The network modularity $M(g)$ of a node partitioning $g$ is equal to the covariance of the ends of a random link of the graph, with distribution $P(i,j)=c_{ij}/2m$, and with respect to the distance $d_g=(1-\delta_{g(i)g(j)})$, as
$$
M(g) = 2\cov_{d_g}(P)
$$
\end{proposition}
\textbf{Proof:} See Appendix \ref{A: modularity and markov stability}. \hfill$\square$
\\
Proposition \ref{propos: correspondence with modularity} shows that the usual interpretation of network modularity is well reproduced in the setting of covariances: for a proposed partitioning $g$, the covariance $\cov_{d_g}(P)$ measures how likely a random link will fall within one of the groups instead of between two different groups. A high covariance thus reflects a good partitioning, in correspondence with the interpretation of a high modularity $M(g)$. In Appendix \ref{A: modularity and markov stability} we show that a generalization of the modularity function, the Markov stability \cite{Delvenne_Markov_stability}, also has a natural interpretation as a covariance.
\section{Maximum variance distributions}\label{S: maxvar distributions main}
We now discuss a number of properties of the variance measured with respect to the square root effective resistance, and consider the problem of finding the largest possible variance on a given graph. It will be useful to consider the variance as a function on the collection of all possible distributions (also called the probability simplex)
$$
\Delta_n \triangleq \left\lbrace \mathbf{p}\in\mathbb{R}^n : p_i\geq 0\text{~and~}\sum_{i=1}^np_i=1 \right\rbrace
$$
as $\var_{\omega}:\Delta_n\rightarrow \mathbb{R}$ defined by $\var_{\omega}(\mathbf{p})=\tfrac{1}{2}\mathbf{p}^T\Omega\mathbf{p}$, where we fix an ordering of the nodes into the columns and rows of the effective resistance matrix $\Omega$. We remark that distributions with $p_i=1$ for a single element $i$ (and thus zero otherwise) correspond to the extreme points of $\Delta_n$, also called the \emph{vertices}. We find the following characterization of the variance function for a certain graph on the probability simplex $\Delta_n$:
\begin{proposition}\label{propos: variance is concave and bounded}
The variance is strictly concave on $\Delta_n$ and bounded by $0\leq \var_{\omega}(\mathbf{p})\leq \sigma^2$. The minimum variance $0$ is attained if and only if the distribution is a vertex of $\Delta_n$, and the maximum variance $\sigma^2$ is attained by a unique distribution $\mathbf{p}^\star$.
\end{proposition}
\textbf{Proof:} Strict concavity of the variance is proven in Appendix \ref{A: variance is concave} based on the definition of the effective resistance in terms of the Laplacian matrix. This concavity means that for any two distinct distributions $\mathbf{p}$ and $\mathbf{q}$, the convex combinations of these distributions $\theta\mathbf{p}+(1-\theta)\mathbf{q}$ for $\theta\in(0,1)$ have a higher variance than the corresponding convex combination of their variances, in other words
$$
\var_\omega(\theta\mathbf{p}+(1-\theta)\mathbf{q})> \theta \var_{\omega}(\mathbf{p}) + (1-\theta)\var_\omega(\mathbf{q}).
$$
For the vertices $\mathbf{p}_i$ of $\Delta_n$ we immediately find that $\var(\mathbf{p}_i)=\omega_{ii}/2=0$. Furthermore, since any other distribution (i.e. not equal to a vertex) can be written as a strict convex combination of vertices, the strict concavity of the variance says that no other distributions can have variance zero, showing that the minimum is only attained for vertices of $\Delta_n$. Next, since all effective resistances and all possible distributions are finite, the variance must be bounded from above by some value $\sigma^2$ which is attained by at least one distribution $\mathbf{p}^\star$. If another distribution $\mathbf{q}^\star$ also were to attain this maximum we would find by strict concavity that $\var_\omega((\mathbf{p}^\star+\mathbf{q}^\star)/2)>\var_\omega(\mathbf{p}^\star)$; since this contradicts the maximality of $\mathbf{p}^\star$, the maximum variance distribution is necessarily unique.\hfill$\square$
\\
Proposition \ref{propos: variance is concave and bounded} thus states that there is a unique solution to the \emph{maximum variance problem}
\begin{align}\label{eq: maximum variance problem}
&\text{maximize~} \frac{1}{2}\mathbf{p}^T\Omega\mathbf{p}\\
&\text{subject to~} \mathbf{p}\in\Delta_n\nonumber.
\end{align}
We write the \emph{maximum variance} as $\sigma^2$, and the \emph{maximum variance distribution} that attains it as $\mathbf{p}^\star$. The support of the maximum variance distribution will be denoted by $\mathcal{V}^\star = \lbrace i:\mathbf{p}^\star_i>0\rbrace$.
\\~\\
\textbf{Numerical solution:}
From Proposition \ref{propos: variance is concave and bounded} we know that the objective function of \eqref{eq: maximum variance problem} is concave. Moreover, the optimization domain $\Delta_n$ is convex which means that the maximum variance problem \eqref{eq: maximum variance problem} on a graph is a \emph{convex quadratic program} and can be solved numerically in a number of steps which is polynomial in the size $n$ of the graph \cite{Boyd_book}. In other words, the maximum variance problem can be solved efficiently. This result was shown before in \cite{Dubuy_Neogy_Solving_quadratic_program} in context of calculating the average weighted resistance distance on a graph. In the Appendix \ref{A: numerical solution of maximum variance distribution}, we discuss how the maximum variance problem can be solved in practice using standard convex programming packages.
\\~\\
\textbf{Analytical solution:} Apart from finding the maximum variance $\sigma^2$ and corresponding distribution $\mathbf{p}^\star$ numerically, we can further characterize this solution in terms of a number of necessary and sufficient conditions for $\mathbf{p}^\star$. As we will show, these conditions allow us to calculate exactly the maximum variance distribution (or its support) for a number of specific graph examples.
\begin{proposition}\label{propos: nec and suff conditions for maxvar distribution}
The maximum variance distribution $\mathbf{p}^\star$ with respect to the square root effective resistance is the unique vector that satisfies the conditions
\begin{align}
&\mathbf{p}\in \Delta_n\textup{with support $\mathcal{V}$}, \tag{$C_1$}\label{eq: 1 nonnegative optimality condition} 
\\
&\Omega_{\mathcal{V}\mathcal{V}}\mathbf{p}_{\mathcal{V}} = 2\var_{\omega}(\mathbf{p})\mathbf{u},\tag{$C_2$}\label{eq: 2 local optimality condition} 
\\
&\tag{$C_3$}\Omega_{\mathcal{V}^c\mathcal{V}}\mathbf{p}_\mathcal{V} < 2\var_{\omega}(\mathbf{p})\mathbf{u}\label{eq: 3 global optimality condition},\\\nonumber
\end{align}
where $(\Omega)_{\mathcal{A}\mathcal{B}}$ denotes a submatrix of $\Omega$ with row (resp. column) indices in the set $\mathcal{A}$ (resp. $\mathcal{B}$), and $(\mathbf{p})_{\mathcal{V}}$ a subvector of $\mathbf{p}$ with indices in $\mathcal{V}$, and with all-one vector $\mathbf{u}=(1,\dots,1)^T$.
\end{proposition}
\textbf{Proof:} Proposition \ref{propos: nec and suff conditions for maxvar distribution} is proven in Appendix \ref{A: proof of necessary and sufficient conditions} in two ways: first, we derive \eqref{eq: 1 nonnegative optimality condition},\eqref{eq: 2 local optimality condition} and \eqref{eq: 3 global optimality condition} as necessary conditions for the maximum variance distribution, and then show that they are also sufficient by concavity of the variance. As a second derivation, we show that the conditions are equivalent to the Karush-Kuhn-Tucker conditions of the optimization problem \eqref{eq: maximum variance problem}.\hfill$\square$
\\
While conditions \eqref{eq: 1 nonnegative optimality condition}$-$\eqref{eq: 3 global optimality condition} might not look very intuitive at first sight, we can interpret them as follows: condition \eqref{eq: 1 nonnegative optimality condition} is a \emph{basic feasibility} criterion which states that $\mathbf{p}^\star$ must be a valid distribution supported on some set of nodes $\mathcal{V}$. Condition \eqref{eq: 2 local optimality condition} is a \emph{local optimality} criterion (local with respect to $\mathcal{V}$) which guarantees that the variance of $\mathbf{p}^\star$ can only decrease if we slightly change its distribution  without changing the support $\mathcal{V}$. Condition \eqref{eq: 3 global optimality condition} finally is a \emph{global optimality} criterion (global w.r.t. $\mathcal{V}$), and guarantees that slightly changing the distribution of $\mathbf{p}^\star$ and changing its support can only result in a decreasing variance. Clearly these are necessary conditions, but as shown in Appendix \ref{A: proof of necessary and sufficient conditions}, they are also sufficient and thus characterize the maximum variance distribution.
\\~\\
\textbf{Maximum variance support $\mathcal{V}^\star$:} Proposition \ref{propos: nec and suff conditions for maxvar distribution} gives an exact description of what the maximum variance distribution looks like for a given graph; more details on the solution to \eqref{eq: 2 local optimality condition} specifically can be found in \ref{S: alternative formulations}. Somewhat surprisingly, the maximum variance distribution can (and often will) have zero probability for a subset of the nodes. Intuitively, this means that when looking for a distribution that is most spread out, some nodes are too \emph{central} in the network (i.e. too close on average to other nodes) and are assigned zero probability as a consequence. The remaining nodes which are in the \emph{maximum variance support} $\mathcal{V}^\star$ on the other hand, are \emph{peripheral} (or boundary) nodes whose distance to other nodes is sufficiently large to contribute to a high variance. By considering the maximum variance distribution, we thus have a procedure to `single out' a subset of nodes $\mathcal{V}^\star$ from a given graph $G$, which can be interpreted as a set of nodes near the boundary or periphery of the graph\footnote{Our use of the term `peripheral' to describe the (geometric) notion of a boundary is different from its use in the context of core-periphery structures, where it is reserved to describe nodes with a certain mesoscale connectivity structure in a graph. While in some cases these might overlap, this is certainly not guaranteed; tree graphs, for instance, can lack a core-periphery structure, but they clearly possess a boundary made up of the leaf nodes.}. In Appendix \ref{A: maximum variance support in some graphs}, we provide further theoretical support for this interpretation by characterizing $\mathcal{V}^\star$ on some simple graphs: trees, weighted stars, configuration graphs, node-transitive graphs and paths, for which the maximum variance support indeed corresponds to the intuitive set of peripheral nodes.
\\~\\
As mentioned in the introduction, our formula for variance also appears in the context of \emph{diversity measures} where it is a special case of Rao's quadratic entropy \cite{Rao} and relates to the R\'{e}nyi entropy of order $2$ when all nodes are assumed indistinguishable (i.e. in a complete graph). In the work of Leinster \emph{et al.} \cite{Cobbold_Leinster, Leinster_book}, these measure were further generalized to a family of diversity measures $\divg_q$ on $\Delta_n$ that encompasses several other well-known measures such as the Gini-Simpson index and our variance fits in this framework\footnote{Instead of describing diversity with respect to distances between points, the framework of Leinster \emph{et al.} starts from similarities between points. As mentioned in \cite{Leinster_book}, one possible mapping is ``similarity = 1-distance", giving rise to a similarity matrix of $\mathbf{u}\mathbf{u}^T-\Omega$ in our case.} as $\divg_2(\mathbf{p}) = (1-2\var(\mathbf{p}))^{-1}$. This relation is particularly relevant in context of Problem \eqref{eq: maximum variance problem}, since it was shown \cite[Thm. 6.3.2]{Leinster_book} that that every similarity space has a unique distribution that \emph{maximizes the diversity measure $\divg_q$ for all $q\in[0,\infty]$ simultaneously}, which thus implies that the maximum variance distribution $\mathbf{p}^\star$ maximizes a whole family of diversity measures.
\subsection{Maximum variance support in random geometric graphs}\label{SS: maximum variance support in RGGs}
\begin{figure}[h!]
    \centering 
    \includegraphics[width = 0.95\textwidth]{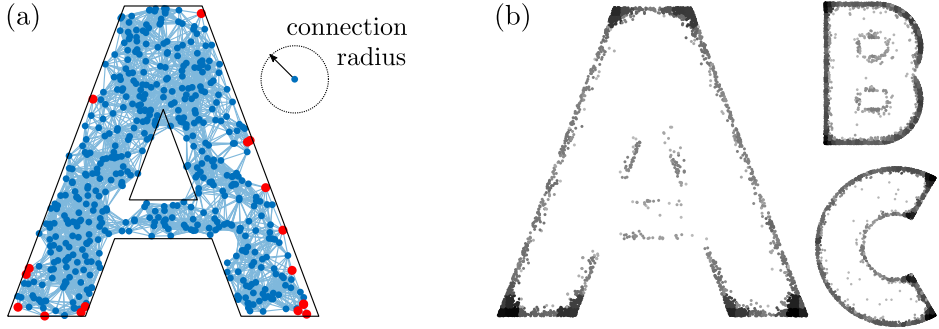}
    \caption{The maximum variance support of random geometric graphs (RGGs) in a bounded domain is more likely to contain nodes which are located near the boundaries. Panel (a) shows one realization of an RGG on $n=500$ nodes in a domain shaped as the letter `A', with connection radius as indicated. The nodes in the maximum variance support are colored red, and are all located near the boundary of the domain. Panel (b) summarizes the maximum variance support node locations calculated in $250$ independent RGG realizations on the given domains. A darker (resp. lighter) shading indicates a higher (lower) density of maximum variance nodes. This figure seems to confirm that the maximum variance support nodes are more likely to be located near the domain boundaries and, in particular, near corners with `high curvature'.}
    \label{fig: boundary on RGGs}
\end{figure}

The results in the previous section and Appendix provide theoretical support for the interpretation that the maximum variance support indicates some sort of periphery or boundary of a graph. Here, we provide another result in this line by considering random geometric graphs (RGGs), which are naturally embedded in some Euclidean space and have a well-defined boundary.
\\
For a set of points $\mathcal{N}\subseteq\mathbb{R}^2$ in the plane, we define the $\epsilon$-graph $G_{\epsilon}$ as a graph with nodes $\mathcal{N}$ and with pairs of nodes linked if and only if they are closer than a certain distance $\epsilon>0$ from each other in the plane (the connection radius). A random geometric graph on a subset $\mathcal{X}\subseteq\mathbb{R}^2$ of the plane is then determined by sampling $n$ points uniformly at random from $\mathcal{X}$ (or via a Poisson point process with a certain rate) and constructing the corresponding $\epsilon$-graph. For more details and more general constructions, see for instance \cite{Penrose_RGGs}. 
\\
In Figure \ref{fig: boundary on RGGs} we show experimentally that the maximum variance support nodes of a random geometric graph on a domain $\mathcal{X}$ tend to be located close to the boundary of this domain. Moreover, this observation seems to hold quite robustly for different parameter settings of the experiment (number of points, connection radius and shape of the domain). We cannot provide a conclusive theoretical explanation of the observations in Figure \ref{fig: boundary on RGGs}, but we believe it could be related to the results of von Luxburg et al. \cite{Luxburg_getting_lost_in_space, vonLuxburg_hitting_and_commute_times}. Their work shows that in certain regimes of $(\epsilon,n)$, the effective resistances in a random geometric graph will be approximately $\omega_{ij}\approx k_i^{-1}+k_j^{-1}$, i.e. as in a configuration graph (see \ref{A: maximum variance support in some graphs}). Consequently, we might conjecture that the maximum variance support in this regime converges accordingly to the result described in Corollary \ref{cor: maximum variance support configuration graph} for configuration graphs in the Appendix, and that the maximum variance support will thus consist of low-degree nodes. As these low-degree nodes are more likely to be located at the boundary of the domain (due to the Poisson-distribution property of Poisson point processes) the maximum variance support nodes would thus indeed be more likely to be close to the domain boundaries. Developing a full theoretical understanding of our observations (potentially following our guesses above), perhaps extending them to RGGs on manifolds (the setting of \cite{vonLuxburg_hitting_and_commute_times}) and further investigating the role of curvature (points near a highly curved boundary will likely have small degrees) seems to be a particularly interesting venue for further research.

\subsection{Application: k-core decomposition from the maximum variance support}\label{SS: new k-core decomposition}
The \emph{$k$-core decomposition} is a network analysis tool which is used to visualise large graphs, identify clusters of important nodes or map the hierarchical structures in a network \cite{Alvarez-Hamelin_k-core_decomposition, Seidman_k-core}. This decomposition divides the nodes of a network into (overlapping) subgraphs by recursively deleting nodes with the smallest degree from a graph until all remaining nodes have at least a certain degree $k$: starting from $k=1$, all nodes are in the $1$-core; then all nodes of degree one are removed consecutively, until all remaining nodes have degree two, which makes up the $2$-core. This process is repeated until no nodes remain.
\\
We now consider an adaptation of the $k$-core decomposition by consecutively removing the boundary nodes, defined as the maximum variance support. This procedure gives a recursive definition of the $k^{\text{th}}$ core $\mathcal{C}_k\subseteq\mathcal{N}$ as
$$
\begin{cases}
\text{$1$-core:~}\mathcal{C}_1 = \mathcal{N}\\
\text{$k$-core:~}\mathcal{C}_k = \mathcal{C}_{k-1}\backslash\mathcal{V}^{\star}(G_{k-1})\text{~for $k>1$}
\end{cases}
$$
where $G_k$ is the induced subgraph on the nodes $\mathcal{C}_{k}$. When a subgraph $G_k$ is disconnected, the maximum variance support is calculated for each connected component separately. The coreness $c_i$ of a node is determined by the `highest' core it is part of, as $c_i=\argmax_k\lbrace i\in\mathcal{C}_k\rbrace$. Figure \ref{fig: k-core figure} below illustrates the $k$-core decomposition on some networks, which shows that our adapted decomposition can fulfill a similar visualisation or summarization role as the standard one.
\begin{figure}[h!]
    \centering
    \includegraphics[width = 0.95\textwidth]{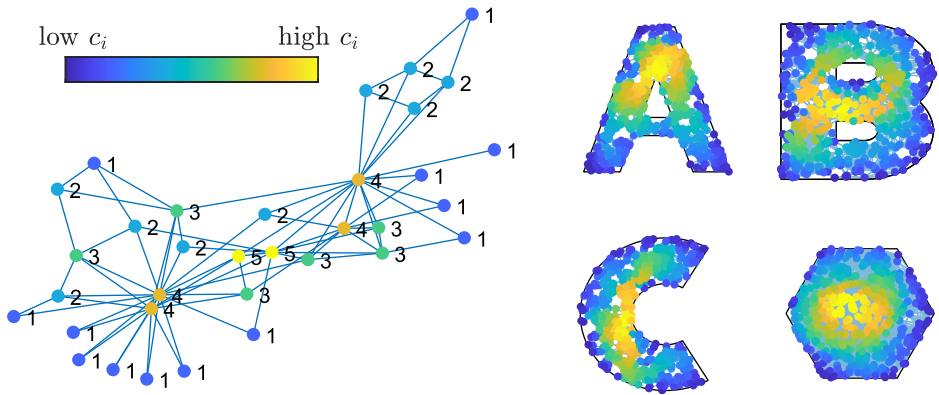}
    \caption{The iterated core-periphery decomposition divides the nodes of a graph into layers of increasing centrality or `coreness'. The figure above shows the coreness $c_i$ of nodes in the Karate club network (obtained from the KONECT database \cite{Kunegis_KONECT}) and four random geometric graphs (see Section \ref{SS: maximum variance support in RGGs}).}
    \label{fig: k-core figure}
\end{figure}
\section{Conclusion}\label{S: conclusion}
In this paper, we have introduced two new measures which allow us to calculate the variance and covariance of distributions defined on the nodes of a network, as a generalization of the standard (co)variance. These measures take into account the underlying structure of the network in the form of distance between nodes, thus providing a tool in the study of functional properties of the network (distributions, signals, ...) relative to the underlying structure of the network. Furthermore, the variance and covariance take the simple form of a quadratic product and matrix trace, respectively, which are easily calculated using standard linear algebra solvers.

To support the specific form of our introduced variance and covariance measures, we show that their definitions specialize to a number of known graph characteristics and heuristics developed in network science and that there is a conceptual correspondence between our (co)variance measures and these graph characteristics.

We furthermore use the variance and covariance in a practical scenario where both structural and functional data of a networked system are known. We analyse a `network of knowledge' consisting of mathematical concepts as nodes and links inferred from hyperlinks on Wikipedia. Based on a corpus of $140$k+ papers on the arXiv, we then analyse how this network of knowledge is used in practice. We translate the occurrences and co-occurrences of concepts in these papers into distributions and joint distributions on the network, and show that the corresponding variance is smaller, on average, compared with a `virtual paper' null model and similarly, that the corresponding covariance is larger, on average, compared with a randomized null model. Beyond this particular setting, our framework has many potential applications in fields like neuroscience, to characterize the relationship between structural and functional brain networks in terms of covariance; in economics, where variance can be used as a measure of economic diversity, or in social networks, where the variance of certain distributions could be interpreted as a measure of polarization. 

In the second part of this paper, we consider the variance on a graph measured with respect to the square root effective resistance distance. In this case, the variance is a strictly concave function over the set of possible distributions $\Delta_n$ and there is a unique maximizing distribution. We give a detailed description of the \emph{maximum variance problem} and the numerical and theoretical approaches to find the maximum variance distribution $\mathbf{p}^\star$. We highlight the interesting observation that, in general, the maximum variance distribution is supported on a subset of the nodes $\mathcal{V}^\star$, which can be interpreted as a set of peripheral/boundary nodes of the underlying graph. This interpretation is supported by analyzing a number of simple graphs such as trees and configuration-like graphs. Additional experimental evidence is found by considering random geometric graphs where we find that the maximum variance support nodes are most likely situated near the boundaries of the domain of the geometric graphs. Following the interpretation of the maximum variance support as a set of peripheral nodes, we propose an application of $\mathcal{V}^\star$ in calculating a $k$-core decomposition of networks.
\\
While the theoretical analyses in this article focus on measuring variance with respect to the effective resistance, we stress that definitions \eqref{eq: definition graph variance} and \eqref{eq: definition graph covariance} work for any distance function on a graph, and in many cases other distances than the resistance distance are likely to be more natural. 

\bibliographystyle{siamplain}
\bibliography{bibliography.bib}

\appendix

\section*{APPENDIX}
\section{Standard variance and covariance as a special case of the graph (co)variance \ref{eq: definition graph variance} and \ref{eq: definition graph covariance}}\label{A: standard (co)variance as special case}
We consider the definition of variance and covariance for distributions on $\mathcal{N}\subset\mathbb{R}$, a finite subset of the real line $\mathbb{R}$, and joint distributions on $\mathcal{N}\times\mathcal{N}$. Both the variance and covariance make use of the \emph{expectation} operator, which is defined as 
\begin{equation}\label{eq: definition expectation}
\mathbb{E}(N) \triangleq \sum_{i\in\mathcal{N}}p(i)i \text{~for some $N\sim p$}.
\end{equation}
This operator can be extended to functions on $\mathcal{N}$ as
$\mathbb{E}(f(N)) = \sum p(i)f(i)$ for some $f:\mathcal{N}\rightarrow\mathbb{R}$, which can be interpreted as the \emph{expected} outcome of the function $f$ when applied to the random variable $N$.
\\
Using the expectation operator, the \emph{variance} of a distribution $p$ is then defined as
\begin{equation}\label{eq: definition standard variance}
\var(p) = \mathbb{E}\left([N-\mathbb{E}(N)]^2\right)\text{~with $N\sim p$}.
\end{equation}
In other words, the variance is the expected squared difference between a random outcome of distribution $p$ and its expected outcome. Importantly, for this definition to make sense we need a `difference' $(i-\mathbb{E}(N))$ between an element of the set and its average, and a way to `square' this difference $(i-\mathbb{E}(N))^2$; both are possible when $\mathcal{N}$ is a subset of the real line.
\\
For a joint distribution $P$, the expectation operator on a function $f:\mathcal{N}\times\mathcal{N}\rightarrow\mathbb{R}$ is defined similarly as $\mathbb{E}(f(N,M))=\sum_{i,j\in\mathcal{N}}P(i,j)f(i,j)$ for a random pair of nodes $(N,M)\sim P$. The \emph{covariance} of a joint distribution $P$ is then defined as
\begin{equation}\label{eq: definition standard covariance}
\cov(P) = \mathbb{E}\left([N-\mathbb{E}(N)][M-\mathbb{E}(M)]\right)\text{~with $(N,M)\sim P$}.
\end{equation}
We will now show that the usual definitions \eqref{eq: definition standard variance} and \eqref{eq: definition standard covariance} for the variance and covariance of distributions on a subset of the real line correspond to the `metric' definitions \eqref{eq: definition graph variance} and \eqref{eq: definition graph covariance}, respectively when taking the Euclidean distance $d^2(i,j)=\Vert i-j\Vert_2^2$.
\\
Introducing expression \eqref{eq: definition expectation} for the (linear) expectation operator into the variance definition \eqref{eq: definition standard variance}, we find
\begin{align*}
\var(p) &= \sum_{i\in\mathcal{N}}p(i)\left[i-\sum_{j\in\mathcal{N}}p(j)j\right]^2
\\
&= \sum_{i\in\mathcal{N}}p(i)\left[i^2 - 2i\sum_{j\in\mathcal{N}}p(j)j + \sum_{j,k\in\mathcal{N}}p(j)p(k)jk\right]
\\
&= \sum_{i,j\in\mathcal{N}}p(i)p(j)\left[i^2 - ij\right]
\\
&= \frac{1}{2}\sum_{i,j\in\mathcal{N}}p(i)p(j)\left[i^2 - 2ij + j^2\right]
\\
&= \frac{1}{2}\sum_{i,j\in\mathcal{N}}p(i)p(j)d^2(i,j)
\end{align*}
as required. For the covariance, we start from the following identities:
$$
\begin{cases}
\cov(P) = \mathbb{E}(NM)-\mathbb{E}(N)\mathbb{E}(M)
\\
\mathbb{E}([N-M]^2) = \mathbb{E}(N^2)-2\mathbb{E}(NM)+\mathbb{E}(M^2)
\end{cases}
$$
which follow from the definition of covariance \eqref{eq: definition standard covariance} and the expectation operator for functions of pairs of random variables. Combining both identities, we find 
\begin{align*}
\cov(P) = \frac{1}{2}\big(&\mathbb{E}(N^2) -2\mathbb{E}(N)\mathbb{E}(M) 
\\&+ \mathbb{E}(M^2) - \mathbb{E}([N-M]^2)\big).
\end{align*}
Expanding the first three terms in this expression, we find
\begin{align*}
&\mathbb{E}(N^2) -2\mathbb{E}(N)\mathbb{E}(M) + \mathbb{E}(M^2) 
\\
&=\sum_{i\in\mathcal{N}}\tilde{p}(i)i^2 -2\sum_{i\in\mathcal{N}}\tilde{p}(i)i\sum_{j\in\mathcal{N}}\tilde{q}(j)j+\sum_{j\in\mathcal{N}}\tilde{q}(j)j^2
\\
&= \sum_{i\in\mathcal{N}}\tilde{p}(i)i\left(i-\sum_{j\in\mathcal{N}}\tilde{q}(j)j\right)+\sum_{j\in\mathcal{N}}\tilde{q}(j)j\left(j-\sum_{i\in\mathcal{N}}\tilde{p}(i)i\right)
\\
&= \sum_{i\in\mathcal{N}}\tilde{p}(i)i\left(\sum_{j\in\mathcal{N}}\tilde{q}(j)(i-j)\right)
\\
&\hphantom{=}+\sum_{j\in\mathcal{N}}\tilde{q}(j)j\left(\sum_{i\in\mathcal{N}}\tilde{p}(i)(j-i)\right)
\\
&= \sum_{i,j\in\mathcal{N}}\tilde{p}(i)\tilde{q}(j)(i-j)^2.
\end{align*}
Consequently, the covariance \eqref{eq: definition standard covariance} can be written as
$$
\cov(P) = \frac{1}{2}\sum_{i,j\in\mathcal{N}}(\tilde{p}(i)\tilde{q}(j)-P(i,j))(i-j)^2
$$
which corresponds to expression \eqref{eq: definition graph covariance} for the Euclidean distance.


\section{Example of joint distributions and their covariance}\label{A: example of covariance}.
Figure \ref{fig: covariance examples} below shows a number of joint distributions on a small graph, together with their covariance and correlation (see Section \ref{S: (co)variance in a network of knowledge}).
\begin{figure}[h!]
    \centering
    \includegraphics[width=0.9\textwidth]{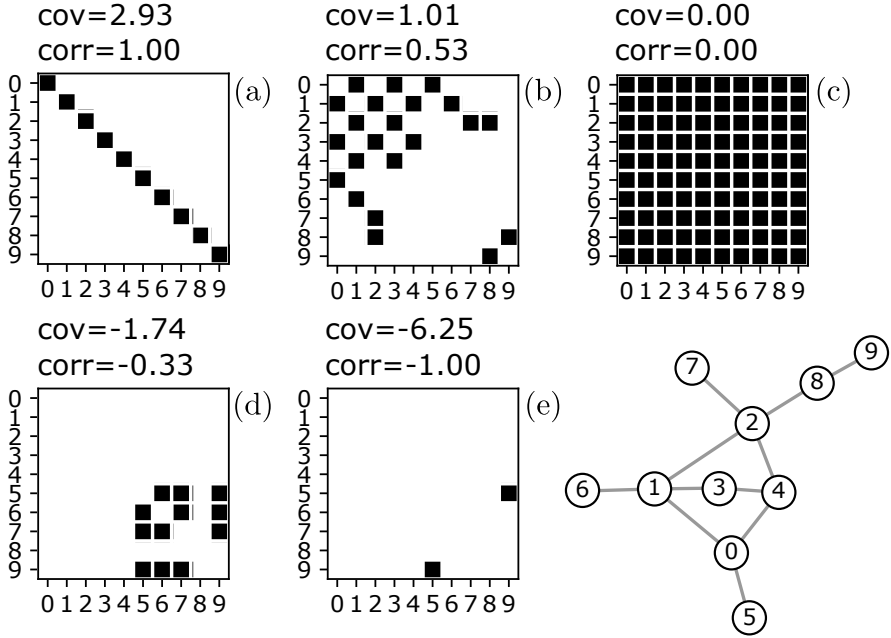}
    \caption{Five joint distributions on a small, 10-node graph. Each joint distribution is illustrated by a $10\times 10$ grid with white squares indicating $P(i,j)=0$ and black squares a constant $P(i,j)=p$ such that $P$ normalizes. The covariance and correlation are calculated with respect to the shortest path distance for (a) the joint distribution for two identical nodes $(N,N)$ with uniform distribution, (b) joint distribution for the random ends of a link, i.e. $(N,M)$ is a link sampled uniformly at random from the graph, (c) the uniform joint distribution, corresponding to two independent uniform random nodes (d) joint distribution for a pair of leaf nodes, i.e. with $(N,M)$ a random (distinct) pair of nodes from $\lbrace 5,6,7,9\rbrace$ and (e) joint distribution for a randomly permuted pair of nodes $(N,M)=(5,9)$ or $(9,5)$.}
    \label{fig: covariance examples}
\end{figure}

\section{Statistical tests for Figure \ref{fig:empirical_variances_compared_to_null_model__and__subfields_variance}}
\label{A: statistics of experiment}
Figure \ref{fig:empirical_variances_compared_to_null_model__and__subfields_variance} in the main text shows a comparison between the variances of concept distributions obtained from real (arXiv) papers, and the variances of concept distributions of virtual papers constructed based on the overall frequencies of the concepts. Visually, the variance distribution of real papers seems to be concentrated on smaller values than the variance distribution of the virtual papers. To quantify this observation, we perform a one-sided Mann-Whitney U test comparing the two distributions (for a given number of concepts in the papers), and the outcome of this test is recorded in Table \ref{table: U test statistic for real and virtual paper comparison}. In all cases, the p-value ($<10^{-12}$) rejects the Mann-Whitney null hypothesis with high significance, in favour of the alternative hypothesis that samples from the null model distribution are typically larger than samples from the empirical distribution i.e. $\Pr[\var(p_{R})<\var(p_{V})]>1/2$ where $p_{R}$ is the concept distribution of a randomly selected real paper, and $p_{V}$ the distribution of a randomly selected virtual paper. In this sense, the variance of the papers is typically smaller than what would be expected from the null model.
\begin{table}[h!]
\begin{center}
\caption{Mann-Whitney U test statistics comparing the empirical distribution of paper variances to the distribution of (null model) virtual paper variances, for papers containing $3-10$ concepts. Tabulated from left to right are the number of concepts, the number of real and virtual papers (no. samples), the U value of the Mann-Whitney test, the normalized U value $U/(n_{\text{real}}n_{\text{virt.}})$ and the $p$-value of the U test statistic.}
\begin{tabularx}{0.99\textwidth}{c|cccc}
no. concepts & no. samples (real/virt.) & U value & U norm. & p-value \\
            \hline\hline
3 concepts & 25716/400 & 6202641.5  & 0.60 & 6.52e-13 \\
4 concepts & 17578/400 & 4995856    & 0.71 &  1.71e-47 \\
5 concepts & 11742/400 & 3553001.5  & 0.76 & 1.13e-68 \\
6 concepts & 7449/400 & 2338463.5   &  0.78 & 1.19e-82 \\
7 concepts & 4545/400 & 1473777.5  &  0.81 & 7.06e-95 \\
8 concepts & 2629/400 & 875333.5   &  0.83 & 2.27e-102 \\
9 concepts & 1428/400 & 478299   &  0.84 & 4.65e-95 \\
10 concepts & 785/400 & 269039  &  0.86 & 2.91e-90
\end{tabularx}
\end{center}
\label{table: U test statistic for real and virtual paper comparison}
\end{table}

\section{Network modularity and Markov stability as covariance}
\label{A: modularity and markov stability}
\textbf{Proof of Proposition \ref{propos: correspondence with modularity}:} For the proposed joint distribution $P(i,j)=\tfrac{c_{ij}}{2m}$ and distance function $d_g(i,j)=1-\delta_{g(i)g(j)}$, we get the covariance expression
$$
\cov_{d_g}(P) = \frac{1}{2}\sum_{i,j\in\mathcal{N}}\left[\frac{k_ik_j}{(2m)^2} - \frac{c_{ij}}{2m}\right](1-\delta_{g(i)g(j)})^2
$$
which by the fact that $c_{ij}=(A)_{ij}$ and $\delta_{g(i)g(j)}^2=\delta_{g(i)g(j)}$ for all $i,j$, and $\sum_{i,j}A_{ij}=\sum_{i,j}k_ik_j/(2m)$ gives equality with the network modularity $M(g)/2$.\hfill$\square$
\\~\\
\textbf{Markov stability and covariance} A recognized problem with the network modularity $M(g)$ is that it suffers from a so-called `resolution limit'. This limit impedes the detection of small but strongly connected groups of nodes \cite{Fortunato_resolution_limit} which is usually undesired when trying to describe the community structure of a network. One method to improve the detection of communities across different scales is the so-called {Markov stability}: the idea is to define the quality of a proposed partitioning $g$ by measuring how likely a (continuous-time) random walker is to remain within the same group after a certain (continuous) time $\tau$. The \emph{Markov stability} $r_\tau(g)$ measures this likelihood and is defined as
$$
r_\tau(g) = \tr\left(T_\tau\operatorname{diag}(\pi)\Delta_g\right)-\pi^T\Delta_g\pi
$$
where $(\Delta_g)_{ij}=\delta_{g(i)g(j)}$ captures the proposed partitioning, $T_\tau$ determines the random walk process as $\Pr(N_{t+\tau}=j\vert N_{t}=i)=(T_\tau)_{ji}$ and with $\pi$ the steady state distribution of this random walk process, i.e. satisfying $T_\tau\pi = \pi$. This transition matrix can be seen as a conditional probability matrix, such that $P_\tau\triangleq T_\tau\operatorname{diag}(\pi)$ is a joint distribution that reflects the probability that the random walker occupies node $i$ in the stationary distribution and that it is at node $j$ after a time $\tau$. In other words distribution $P_\tau$ describes the distribution of a pair of nodes $(N_{t},N_{t+\tau})$. Again, the formula for the Markov stability is reminiscent of the graph covariance \eqref{eq: definition graph covariance} and we find the following correspondence:
\begin{proposition}\label{propos: correspondence with Markov stability}
The Markov stability $r_\tau(g)$ of a node partitioning $g$ is equal to the covariance of the position of a stationary random walker and its position after time $\tau$, with respect to the distance $d_g$, as
$$
r_\tau(g) = 2\cov_{d_g}(P_\tau)
$$
\end{proposition}
\textbf{Proof:} The proposed covariance can be written as
\begin{align*}
\cov_{d_g}(P_{\mathcal{T}}) &= \frac{1}{2}\sum_{i,j\in\mathcal{N}}\left(\pi(i)\pi(j) - (T_\tau)_{ij}\pi(i)\right)\left(1-\delta_{g(i)g(j)}\right)^2
\end{align*}
which by the fact that $\sum_{i,j}\pi(i)\pi(j)=\sum_{i,j}(T_\tau)_{ij}\pi(i)$ gives equality with the Markov stability $r_\tau(g)/2$.\hfill$\square$
\\
In \cite{Delvenne_Markov_stability} the Markov stability measure is derived based on a covariance matrix of the random vectors $\mathbf{X}(t)=(X_1(t),\dots,X_n(t))^T$ and $\mathbf{X}(t+\tau)$, where $X_i(t)$ is a Bernoulli random variable, with probability $\Pr(X_i(t)=1)=\Pr(N_t=i)$. The Markov stability is then found as the trace of this covariance matrix, after taking into account the proposed partitioning. By Proposition \ref{propos: correspondence with Markov stability}, this methodology is thus equivalent to simply measuring the covariance of the joint distribution (in contrast with the covariance of two random vectors) with respect to the metric $d_g$. 
\\~\\
In Appendix \ref{A: variance for diffusion} we briefly consider the variance for diffusion processes on a graph, which are closely related to the continuous-time random walk process used in the Markov stability.

\section{Strict concavity of the variance}\label{A: variance is concave}
A function $f$ on a set $\mathcal{X}$ is called \emph{strictly convex} if it satisfies
$$
f(\theta x+(1-\theta)y) > \theta f(x) + (1-\theta)f(y)
$$
for all distinct $x,y\in\mathcal{X}$ and $\theta\in(0,1)$. We now prove the following Lemma:
\begin{lemma}
The variance $\var_{\omega}$ is strictly convex on $\Delta_n$.
\end{lemma}
\noindent\textbf{Proof:} For any $\theta\in[0,1]$ and $\mathbf{p},\mathbf{q}\in\Delta_n$ we can write
\begin{align*}
&\theta\mathbf{p}^T\Omega\mathbf{p} + (1-\theta)\mathbf{q}^T\Omega\mathbf{q} - [\theta\mathbf{p}+(1-\theta)\mathbf{q}]^T\Omega[\theta\mathbf{p}+(1-\theta)\mathbf{q}]\\
&=\theta(1-\theta)\left[\mathbf{p}^T\Omega\mathbf{p}+ \mathbf{q}^T\Omega\mathbf{q} - 2\mathbf{p}^T\Omega\mathbf{q}\right]\\
&=\theta(1-\theta)(\mathbf{p}-\mathbf{q})^T\Omega(\mathbf{p}-\mathbf{q})\\
&= -2\theta(1-\theta)(\mathbf{p}-\mathbf{q})^TQ^\dagger(\mathbf{p}-\mathbf{q}) \leq 0.
\end{align*}
In the last line, we first use the fact that by definition of the effective resistance we have $\Omega=\mathbf{u}\zeta^T+\zeta \mathbf{u}^T-2Q^\dagger$ where $\zeta=\operatorname{diag}(Q^\dagger)$, such that $\mathbf{x}^T\Omega\mathbf{x} = -2\mathbf{x}^TQ^\dagger\mathbf{x}$ for all vectors $\mathbf{x}\perp \mathbf{u}$, and thus in particular for $(\mathbf{p}-\mathbf{q})$. Since the Laplacian matrix is positive semidefinite (see for instance \cite{karel_resistance>distance}), its pseudoinverse $Q^\dagger$ is positive semidefinite as well, resulting in the inequality and thus establishing concavity of the variance. Moreover, since the (pseudoinverse) Laplacian has a single zero eigenvalue corresponding to the constant eigenvector \cite{karel_resistance>distance} and since $(\mathbf{p}-\mathbf{q})\perp \mathbf{u}$ for distinct distributions, the inequality is strict.\hfill$\square$
\\
We remark that the concavity of the variance is not a generic property for all metrics $d$. For some metric on the nodes $d$ with matrix $(D)_{ij}=d^2(i,j)$ we can write
\begin{align*}
\theta&\var_d(\mathbf{p})+(1-\theta)\var_d(\mathbf{p}) -\var_d(\theta\mathbf{p}+(1-\theta)\mathbf{q})\\
&=\theta(1-\theta)(\mathbf{p}-\mathbf{q})^T D(\mathbf{p}-\mathbf{q}).
\end{align*}
This shows that the variance $\var_d$ is concave if and only if 
\begin{equation}\label{eq: concavity criterion}
(\mathbf{p}-\mathbf{q})^T D(\mathbf{p}-\mathbf{q})\leq 0\text{~for all~}\mathbf{p}\neq\mathbf{q}\in\Delta_n.
\end{equation}
A result from \emph{distance geometry} states that criterion \eqref{eq: concavity criterion} together with the conditions $d(i,i)=0$ and $d(i,j)=d(j,i)$ for all $i,j$ is satisfied if and only if the metric space $(\mathcal{N},\sqrt{d})$ can be embedded isometrically in Euclidean space \cite{Blumenthal_distance_geometry}, i.e. with a point $\mathbf{m}(i)\in\mathbb{R}^r$ for each $i$ such that $\sqrt{d(i,j)}=\Vert\mathbf{m}(i)-\mathbf{m}(j)\Vert^2$. In Section \ref{S: alternative formulations} we discuss how this `embedding criterion' is satisfied for the effective resistance as discovered by Miroslav Fiedler \cite[Thm. 1.2.4]{Fiedler_matrices_and_graphs}. Similar criteria for distance matrices were also studied in \cite{Hjorth_finite_metric_spaces}.


\section{Proof of Proposition \ref{propos: nec and suff conditions for maxvar distribution} on maximum variance distributions}
\label{A: proof of necessary and sufficient conditions}
To further characterize the solution to the maximum variance problem we will first consider some necessary conditions that must be satisfied by any candidate optimal solution. In what follows, we assume that we know the maximum variance distribution $\mathbf{p}$ and derive three necessary conditions for this distribution. We then show that these conditions are also sufficient in the case of $\var_{\omega}$, and thus fully characterize the maximum variance distribution that solves problem \eqref{eq: maximum variance problem}.
\\
A first necessary condition is a simple consequence of the fact that $\mathbf{p}$ is a distribution and thus needs to have nonnegative entries that sum to one; in other words it must satisfy the (entrywise) inequality and equality
\begin{equation}\label{eq: 1S nonnegative optimality condition}
\mathbf{p}\geq 0\text{~and~}\mathbf{u}^T\mathbf{p}=1.
\end{equation}
The non-negativity condition may be satisfied with equality for some nodes (i.e. $p(i)=0$ for some $i$), which can be captured in the \emph{support} of the distribution as $\supp(\mathbf{p})=\lbrace i\in\mathcal{N}:p(i)>0\rbrace$. We will further abbreviate the support of our tentative optimal solution $\mathbf{p}$ by the set $\mathcal{V}\subseteq\mathcal{N}$ of size $v$.
\\
Further optimality criteria can be derived from the fact that the maximal distribution $\mathbf{p}$ has the largest variance amongst all possible distributions, which means that any small perturbation of this distribution to another distribution must result in a decreasing variance. A first type of perturbation is a small transfer of probability mass $\epsilon$ between two nodes of the support $i,j\in\mathcal{V}$, which yields the perturbed distribution $\mathbf{p}'=\mathbf{p}-\epsilon(e_i-e_j)$. Comparing the variance of this distribution to the original variance, we then must have
$$
\var_{\omega}(\mathbf{p}')-\var_{\omega}(\mathbf{p})\leq 0\text{~for all $i,j\in\mathcal{V}$ and $\epsilon$}.
$$
Introducing expression \ref{eq: definition resistance variance and covariance} for the variance, this condition translates to the inequality $\epsilon(e_i-e_j)^T\Omega\mathbf{p}\geq -\epsilon^2\omega_{ij}$. Since this must be true for all pairs of nodes in the support and all (small) values of $\epsilon$, we find the necessary condition\footnote{Since $\omega_{ij}\geq 0$ and $\epsilon$ can be both positive and negative, we have that $\vert(e_i-e_j)\Omega\mathbf{p}\vert\leq \vert \epsilon\omega_{ij}\vert$.}
\begin{equation}\label{eq: 2S local optimality condition}
    \Omega_{\mathcal{V}\mathcal{V}}\mathbf{p}_{\mathcal{V}} = \alpha \mathbf{u}\text{~for some $\alpha\in\mathbb{R}$},
\end{equation}
where $\Omega_{\mathcal{V}\mathcal{V}}$ is the $v\times v$ principal submatrix of the effective resistance matrix $\Omega$, containing the resistances between pairs of nodes in the support, and similarly the $v\times 1$ vector $\mathbf{p}_\mathcal{V}$ containing the probabilities of nodes in the support. Equation \eqref{eq: 2S local optimality condition} is the second necessary condition for the maximum variance distribution, and it is a \emph{local optimality criterion} in the sense that it characterizes the optimal solution amongst all solutions with a given support $\mathcal{V}$. In Section \ref{S: alternative formulations}, we show that equation \eqref{eq: 2S local optimality condition} has a unique solution that is consistent with the normalization condition $\mathbf{u}^T\mathbf{p}=1$, which can be encoded by letting\footnote{This value for $\alpha$ follows immediately by multiplying both sides of \eqref{eq: 2S local optimality condition} by $\mathbf{p}^T$.} $\alpha=2\var_{\omega}(\mathbf{p})$.
\\
The solution to equation \eqref{eq: 2 local optimality condition} is thus a candidate maximum variance distribution \emph{if} this solution is a distribution. The question then remains which of these tentative distributions is the global optimum (i.e. for which $\mathcal{V}$). For this reason, we consider a second type of perturbation which amounts to a small transfer of probability mass $\epsilon>0$ from a node in the support $i\in\mathcal{V}$ to a node outside of the support $j\in\mathcal{V}^c$, yielding the perturbed distribution $\mathbf{p}''=\mathbf{p}-\epsilon(e_i-e_j)$. Comparing the variance of this distribution to the original variance, we thus must have
$$
\var_{\omega}(\mathbf{p}'')-\var_{\omega}(\mathbf{p})\leq 0\text{~for all $i\in\mathcal{V},j\in\mathcal{V}^c$ and $\epsilon\geq 0$},
$$
which translates again to inequality $2(e_i-e_j)^T\Omega\mathbf{p}\geq 0$ when introducing the variance and letting $\epsilon\rightarrow 0+$. From the first two necessary conditions, we furthermore know that $e_i^T\Omega\mathbf{p}=2\var_\omega(\mathbf{p})$ for any $i\in\mathcal{V}$, such that these inequalities can be written as the following (entrywise) inequality
\begin{equation}\label{eq: 3S global optimality condition}
\Omega_{\mathcal{V}^c\mathcal{V}}\mathbf{p}_\mathcal{V} \leq 2\var_{\omega}(\mathbf{p})\mathbf{u},
\end{equation}
where $\Omega_{\mathcal{V}^c\mathcal{V}}$ is the $(n-v)\times v$ submatrix of the effective resistance matrix $\Omega$, containing columns in $\mathcal{V}$ and rows in its complement $\mathcal{V}^c$. Equation \eqref{eq: 3S global optimality condition} is the third necessary condition for the maximum variance distribution, and it is a \emph{global optimality criterion} since it guarantees that the variance decreases when changing the support of $\mathbf{p}$ by transferring probability mass to nodes outside the support.
\\
To summarize, we have identified three necessary conditions for the maximum variance distribution $\mathbf{p}^\star$ that solves problem \eqref{eq: maximum variance problem} for \emph{any} distance function; condition \eqref{eq: 1S nonnegative optimality condition} guarantees that our solution is a distribution and condition \eqref{eq: 2S local optimality condition} and \eqref{eq: 3S global optimality condition} make sure that any local change in the distribution leads to a decrease in variance. In other words, these three conditions characterize a \emph{local maximum} of the variance in the domain,
Now, since the variance with respect to the resistance distance is concave this local solution also corresponds to the \emph{global maximum}, i.e. the necessary conditions are also sufficient, and we find the following characterization:
\begin{proposition*}[Repeated from main text]
The maximum variance distribution $\mathbf{p}^\star$ with respect to the square root effective resistance is the unique vector that satisfies the conditions
\begin{align}  
&\mathbf{p}\in \Delta_n\textup{with support $\mathcal{V}$}, \tag{$C_1$}\label{eq: 1SS nonnegative optimality condition} 
\\ 
&\Omega_{\mathcal{V}\mathcal{V}}\mathbf{p}_{\mathcal{V}} = 2\var_{\omega}(\mathbf{p})\mathbf{u},\tag{$C_2$}\label{eq: 2SS local optimality condition} 
\\ 
&\Omega_{\mathcal{V}^c\mathcal{V}}\mathbf{p}_\mathcal{V} \leq  2\var_{\omega}(\mathbf{p})\mathbf{u}\tag{$C_3$}\label{eq: 3SS global optimality condition},\\\nonumber
\end{align}
for some nonempty set $\mathcal{V}\subseteq\mathcal{N}$, and where $(\Omega)_{\mathcal{A}\mathcal{B}}$ denotes a submatrix of $\Omega$ with row (resp. column) indices in the set $\mathcal{A}$ (resp. $\mathcal{B}$), and $(\mathbf{p})_{\mathcal{V}}$ a subvector of $\mathbf{p}$ with indices in $\mathcal{V}$.
\end{proposition*}
\textbf{Proof:} From the derivation above, we know that the three conditions are necessary. It thus remains to show that they are sufficient.\\
As the variance is a strictly concave function, we can bound the variance of any distribution $\mathbf{q}\neq\mathbf{p}$ as
$$
\var_{\omega}(\mathbf{q}) < \var_{\omega}(\mathbf{p}) + (\mathbf{q}-\mathbf{p})^T\nabla\var_{\omega}(\mathbf{p})
$$
where the gradient of the variance equals $\nabla\var_{\omega}(\mathbf{p})=\Omega\mathbf{p}$. Rewriting the difference between distributions we find that
$$
\var_{\omega}(\mathbf{q})< \var_{\omega}(\mathbf{p}) + \sum_{i\in\mathcal{V},j\in\mathcal{N}}p_iq_j(e_j-e_i)^T\Omega\mathbf{p}
$$
When a distribution $\mathbf{p}$ satisfies conditions \eqref{eq: 1 nonnegative optimality condition}$-$\eqref{eq: 3 global optimality condition}, the terms in the sum are all non-positive and thus $\var_{\omega}(\mathbf{q}) < \var_{\omega}(\mathbf{p})$ holds for all $\mathbf{q}\in\Delta_n$ different from $\mathbf{p}$. This shows that these conditions are not only necessary, but also sufficient for the vector $\mathbf{p}$ to be the global optimum $\mathbf{p}^\star$, which proves the proposition.\hfill$\square$
\\
\textbf{Alternative proof:} From the main text, we recall that the maximum variance problem is a convex quadratic problem\footnote{The convex/concave terminology might be confusing, but this refers to the standard setup of optimization problems as finding the \emph{minimum} of a \emph{convex problem}, instead of finding the maximum of a concave problem as in the maximum variance problem. With a simple change of sign, our problem belongs to the class of convex quadratic (minimization) problems.} \cite{Boyd_book} which asks to maximize a concave objective function $\var_{\omega}(\mathbf{p})$ subject to convex inequality constraints $p_i\geq 0$ and linear equality constraints $\mathbf{u}^T\mathbf{p}=1$. Furthermore, there exists points that satisfy all inequality conditions strictly as $p_i>0$ (take for instance $\mathbf{u}/n$), which means that the maximum variance problem satisfies \emph{Slater's condition} \cite[Ch. 5.2.3]{Boyd_book}.
\\
A convex quadratic problem that satisfies Slater's condition can be solved by defining the Lagrangian $L$ associated to the problem, as
$$
L(\mathbf{p},\boldsymbol{\lambda},\nu) = \tfrac{1}{2}\mathbf{p}^T\Omega\mathbf{p} + \boldsymbol{\lambda}^T\mathbf{p} + \nu(\mathbf{u}^T\mathbf{p}-1),
$$
where $\boldsymbol{\lambda}\in\mathbb{R}^n$ and $\nu\in\mathbb{R}$ are called the \emph{Lagrangian multipliers} associated with the constraints on $\mathbf{p}$. The solution of the problem is then found from the so-called Karush-Kuhn-Tucker (KKT) conditions \cite[Ch. 5.5.3]{Boyd_book}:
$$
\begin{dcases}
\nabla_{\mathbf{p}}L = 0\\
\mathbf{u}^T\mathbf{p} = 1\\
p_i\geq 0 \text{~for $i=1,\dots,n$}\\
\lambda_i\geq 0 \text{~for $i=1,\dots,n$}\\
\lambda_ip_i = 0\text{~for $i=1,\dots,n$}
\end{dcases}
$$
where $\nabla_{\mathbf{p}}$ is the gradient with respect to $\mathbf{p}$. We now show that these KKT conditions for $\mathbf{p},\boldsymbol{\lambda},\nu$ are equivalent to the conditions for $\mathbf{p}$ in Proposition \ref{propos: nec and suff conditions for maxvar distribution}. First, making the gradient explicit and introducing the set $\mathcal{V}$, the KKT conditions can be written equivalently as
$$
\begin{cases}
\Omega\mathbf{p}+\boldsymbol{\lambda}+\nu\mathbf{u}=0\\
\mathbf{u}^T\mathbf{p}=1\\
p_i>0\text{~for $i\in\mathcal{V}$}\\
p_i=0\text{~for $i\in\mathcal{V}^c$}\\
\lambda_i= 0\text{~for $i\in\mathcal{V}$}\\
\lambda_i\geq 0\text{~for $i\in\mathcal{V}^c$}\\
\end{cases}
\Leftrightarrow
\begin{cases}
\Omega\mathbf{p} + \boldsymbol{\lambda} + \nu\mathbf{u}=0\\
\mathbf{p}\in\Delta_n\text{~with supp. $\mathcal{V}$}\\
\lambda_i=0\text{~for $i\in\mathcal{V}$}\\
\lambda_{i}\geq0\text{~for $i\in\mathcal{V}^c$}
\end{cases}
\Leftrightarrow
\begin{cases}
\Omega_{\mathcal{V}\mathcal{V}}\mathbf{p}_{\mathcal{V}} = -\nu\mathbf{u}\\
\Omega_{\mathcal{V}^c\mathcal{V}}\mathbf{p}_{\mathcal{V}} = -\nu\mathbf{u} - \boldsymbol{\lambda}_{\mathcal{V}^c}\\
\mathbf{p}\in\Delta_n\text{~with supp. $\mathcal{V}$}.
\end{cases}
$$
Going from the first set of conditions to the second, the conditions on $\mathbf{p}$ are combined. From the second set of conditions to the third, the zero-gradient condition is split up into the equations corresponding to $\mathcal{V}$ and those corresponding to $\mathcal{V}^c$. From this final set of conditions, we can use the normalization $\mathbf{u}^T\mathbf{p}=1$ to find that $\nu=-2\var_{\omega}(\mathbf{p})$ and the inequality $\boldsymbol{\lambda}_{\mathcal{V}^c}\geq 0$ to arrive at the conditions \eqref{eq: 1 nonnegative optimality condition},\eqref{eq: 2 local optimality condition} and \eqref{eq: 3 global optimality condition} of Proposition \ref{propos: nec and suff conditions for maxvar distribution}. Since the KKT conditions are necessary and sufficient conditions for a solution to the maximum variance problem, this concludes the proof. \hfill$\square$

\section{Maximum variance support on some particular graphs}\label{A: maximum variance support in some graphs}
One of the interesting features of the maximum variance distribution $\mathbf{p}^\star$ is that it can be supported on a (strict) subset $\mathcal{V}^\star$ of the nodes. As discussed in the main text, we may in that case think of the maximum variance support as indicating a set of ``boundary nodes'' in the network. Here, we support this intuition by showing that for a number of simple graphs, the maximum variance support indeed corresponds to peripheral or boundary nodes in the graph. These results rely on an exact solution of $\mathbf{p}^\star$ and $\mathcal{V}^\star$ from the conditions described in Proposition \ref{propos: nec and suff conditions for maxvar distribution}.\\
\textbf{Tree graphs:} A connected graph $G$ is called a \emph{tree graph} if there is exactly one path between every two nodes, or equivalently, if it contains no cycles. In many ways, tree graphs are the simplest possible graphs and their properties can be studied in great detail. The nodes of a tree graph fall in two categories: the \emph{leaf nodes}\footnote{A standard result of graph theory states that every tree has at least $2$ leaf nodes; see for instance \cite{Bollobas_graph_theory}} which have a single incident link and form the `extremities' of the tree, and the \emph{non-leaf nodes} that make up the `core' of the tree; every path between two leaf nodes necessarily passes through one of the non-leaf nodes. This intuitively clear distinction between central and peripheral nodes in trees is well reproduced when considering the maximum variance support $\mathcal{V}^\star$ on tree graphs; we find the following condition:
\begin{proposition}\label{propos: maxvar support on trees}
On tree graphs, the maximum variance support $\mathcal{V}^\star$ is a subset of the leaf nodes.
\end{proposition}
\textbf{Proof:} A proof of Proposition \ref{propos: maxvar support on trees} was presented in \cite{Dankelman_average_distance} when considering the equivalent problem of maximizing $\mathbf{p}^T D\mathbf{p}$ for shortest-path distance matrix $D$ (which equals the effective resistance matrix $\Omega$ in the case of tree graphs) over all distributions $\mathbf{p}\in\Delta_n$. For completeness, we include a self-contained proof in Appendix \ref{A: maxvar for tree graphs} based on our expressions for the maximum variance distribution. Moreover, the proof and thus Proposition \ref{propos: maxvar support on trees} is valid for any distance function $d$ for which $d(i,j)=d(i,x)+d(x,j)$ whenever removing $x$ from the graph disconnects $i$ and $j$.\hfill$\square$
\\
~
\\
\textbf{Weighted star graphs}: As discovered already in \cite{Hjorth_finite_metric_spaces, Dankelman_average_distance}, the maximum variance support on a tree can be a strict subset of the leaf nodes in some cases. To illustrate this, we further restrict our attention to (weighted) star graphs as a special case of tree graphs. A star graph on $n+1$ nodes consists of a central (non-leaf) node $\lbrace 0\rbrace$ connected to all the other (leaf) nodes, with link weights $c_{0i}=k_i$, and no further connections otherwise. The leaf node degrees $\mathbf{k}=\lbrace k_1,\dots,k_n\rbrace$ fully parametrize a weighted star and all relevant properties can be expressed in terms of these degrees. In particular, the effective resistance between any two leaf nodes $i$ and $j$ is given by $\omega_{ij}=k_i^{-1}+k_j^{-1}$ as shown in Appendix \ref{A: maxvar for stars and config graphs}. For star graphs, we then find the following exact characterization of the maximum variance support:
\begin{proposition}\label{prop: maximum variance support weighted star}
On a weighted star graph with leaf-node degrees $\mathbf{k}$, the maximum variance support $\mathcal{V}^\star$ is equal to the nodes with the $\ell$ smallest degrees, such that $k_\ell\leq (\ell-2)^{-1}\sum_{i=1}^\ell k_i< k_{\ell+1}$.
\end{proposition}
\textbf{Proof:} We prove Proposition \ref{prop: maximum variance support weighted star} in Appendix \ref{A: maxvar for stars and config graphs} based on the necessary conditions \eqref{eq: 1 nonnegative optimality condition}$-$\eqref{eq: 3 global optimality condition} for the maximum variance distribution.\hfill$\square$
\\
Using Proposition \ref{prop: maximum variance support weighted star} we find that the weighted star with degrees $\lbrace k,k,k,k'\rbrace$ is supported on the nodes with degree $k$, i.e. a strict subset of the leaf nodes, whenever $k'>3k$, corresponding to an example given in \cite{Hjorth_finite_metric_spaces}. 
\\~\\
\textbf{Configuration graphs}: While very simple and somewhat artificial, the problem of finding the maximum variance support on a star graph is in fact equivalent to the same problem on a more popular type of graph.
For a (multi)set of degrees $\mathbf{k}=\lbrace k_1,\dots,k_n\rbrace$, the \emph{configuration graph} $C_\mathbf{k}$ is a graph on $n$ nodes where each node $i$ is assigned one of the degrees $k_i$ and has link weights $c_{ij}=k_ik_j/(2m-1)$ to all other nodes $j$. These graphs appear as the ensemble average\footnote{More precisely, for the configuration model which allows for multiple links and self loops, the probability that two nodes $i,j$ are connected by a link equals $k_ik_j/(2m-1)$. This includes a probability of $k_i(k_i-1)/4m$ that a node has a self-loop, which can not be captured using our Laplacian representation.} of the popular \emph{configuration random graphs} (see e.g. \cite{Newman_Networks}) which are frequently used to study the influence of the degree sequence on various graph properties, or as null models in the analysis of empirical graphs; configuration graphs also appear in the mean-field analysis of dynamical processes on graphs such as epidemics \cite{PastorSatorras_epidemics}. In Appendix \ref{A: maxvar for stars and config graphs} we show that the effective resistance between any two nodes in $C_{\mathbf{k}}$ equals $\omega_{ij}=(2m-1)/(2m)(k_i^{-1}+k_j^{-1})$ as in the star graph, which means that solving problem \eqref{eq: maximum variance problem} on configuration graphs and star graphs with the same degree sequence $\mathbf{k}$ is equivalent, up to some constant factor. We thus have the following Corollary of Proposition \ref{prop: maximum variance support weighted star}
\begin{corollary}\label{cor: maximum variance support configuration graph}
On configuration graphs with degree sequence $\mathbf{k}$, the maximum variance support $\mathcal{V}^\star_{\mathbf{k}}$ equals the nodes with the $\ell$ smallest degrees, where $
k_\ell\leq(\ell-2)^{-1}\sum_{i=1}^\ell k_i< k_{\ell+1}.
$
\end{corollary}
\textbf{Proof:} The proof of Corollary \ref{cor: maximum variance support configuration graph} is given in Appendix \ref{A: maxvar for stars and config graphs} by establishing the equivalence of the maximum variance problem \eqref{eq: maximum variance problem} on weighted stars and configuration graphs with the same degree sequence $\mathbf{k}$.\hfill$\square$
\\
It was shown in \cite{Luxburg_getting_lost_in_space, vonLuxburg_hitting_and_commute_times} that in certain parameter regimes of random geometric graphs and for realizations of the configuration model, the effective resistance between all node pairs will approximately be proportional to $k_i^{-1}+k_j^{-1}$, in which case $\mathcal{V}^\star_{\mathbf{k}}$ is thus particularly suitable as an approximate maximum variance support. Furthermore, this result could explain our observation in Section \ref{SS: maximum variance support in RGGs} that nodes in the maximum variance support of a random geometric graphs seem more likely to be located close to the boundaries of the sampled domain.
\\
~
\\
\textbf{Node-transitive graphs}: Our results above focus on the case where the maximum variance support $\mathcal{V}^\star$ points to a subset of the total node set. In some cases however, a graph might be `homogeneous' in the sense that all nodes have similar distances to the other nodes and no central or peripheral node sets stand out. In the example above for instance, when the degree sequence satisfies $k_{\max}/k_{\min}<v/(v-2)$, all nodes will be sufficiently similar with $\mathcal{V}^\star=\mathcal{N}$ as a result. An example of such homogeneous graphs are \emph{node-transitive graphs} (more commonly called vertex-transitive graphs). A graph $G=(\mathcal{N},\mathcal{L})$ is node transitive \cite{Godsil_royle_algebraic_graph_theory} if for every pair of nodes $i$ and $j$, there exists a bijection $\pi:\mathcal{N}\rightarrow\mathcal{N}$ that maps $i$ to $j$ as $\pi(i)=\pi(j)$ and where $\pi$ is an automorphism such that $(\pi(a),\pi(b))\in\mathcal{L}$ if and only if $(a,b)\in\mathcal{L}$. In other words, no two nodes (and thus no subset of nodes) are `distinguishable' from each other by simply considering their position in the network. Since all nodes in the network are indistinguishable, the corresponding Laplacian and resistance matrices are highly symmetrical from which the following result follows:
\begin{proposition}\label{prop: maxvar on node transitive}
On node-transitive graphs, the maximum variance distribution is the uniform distribution $\mathbf{p}^\star=\mathbf{u}/n$ and the maximum variance support $\mathcal{V}^\star$ is the full node set $\mathcal{N}$.
\end{proposition}
\textbf{Proof:} In \cite[Thm. 14]{Zhou_on_resistance_matrix_node-transitive} it is shown that $\sum_{j\sim i}c_{ij}\omega_{ij}=2-2/n$ for any node $i$ in a node-transitive graph. Following Proposition \ref{propos: equivalent characterizations} (see later) we thus find
$p^\star_i=1/n$ for every node and consequently $\mathcal{V}^\star=\mathcal{N}$.\hfill$\square$
\\
Some well-known examples of node-transitive graphs to which this result thus applies are complete graphs, cycles and hypercubes as well as all Cayley graphs (of which the former are all examples) \cite{Godsil_royle_algebraic_graph_theory}. We remark that a maximum variance distribution supported on the full node set has strong `topological' implications for the underlying graph; Fiedler found that in this case, removing any set of $r$ nodes can disconnect the graph into at most $r-1$ components and conjectured that, conversely, any graph with this connectivity property must have $\mathcal{V}^\star=\mathcal{N}$ \cite[Thm. 3.4.18]{Fiedler_matrices_and_graphs}.
\\
~
\\
\textbf{Path graphs:} To conclude, we observe that the maximum variance support only contains two nodes in the case of a path graph, more precisely we find:
\begin{proposition}
The maximum variance support contains exactly two nodes if and only if these nodes are the ends of a path graph.
\end{proposition}
\textbf{Proof:} For a set of two nodes $\mathcal{V}=\lbrace a,b\rbrace$ we find that equation \eqref{eq: 2 local optimality condition} yields $p_a=p_b=1/2$ and that
\begin{align*}
    \sum_{i\in\mathcal{V}}\omega_{ix}p_i - \sum_{i,j\in\mathcal{V}}\omega_{ik}p_ip_k = \frac{1}{2}(\omega_{ax}+\omega_{xb}-\omega_{ab})\geq 0
\end{align*}
for all $x\notin\mathcal{V}$. This shows that condition \eqref{eq: 3 global optimality condition} is satisfied if and only if $\omega_{ax}+\omega_{xb}=\omega_{ab}$ for all $x\notin\mathcal{V}$, which is equivalent to $a$ and $b$ being the ends of a path graph.\hfill$\square$
\\
We remark that this contradicts an example proposed in \cite{Dankelman_average_distance} about a three-leaf tree graph with maximum variance support on two leaf nodes.

\section{Calculating maximum variance distribution}\label{A: numerical solution of maximum variance distribution} Following its characterization as a convex quadratic program, we know that the maximum variance distribution can be calculated efficiently. In practice, this can be done using existing convex programming software -- on our GitHub page \cite{github}, we provide example implementations for MATLAB and Python. In Figure \ref{fig: degrees and maxvar distributions} below, we show the result for a number of empirical graphs, calculated using the CVX package \cite{cvx, Grant_boyd_cvx}:
\begin{figure}[h!]
    \centering
    \includegraphics[scale=0.8,trim=1cm 0 0 0,clip]{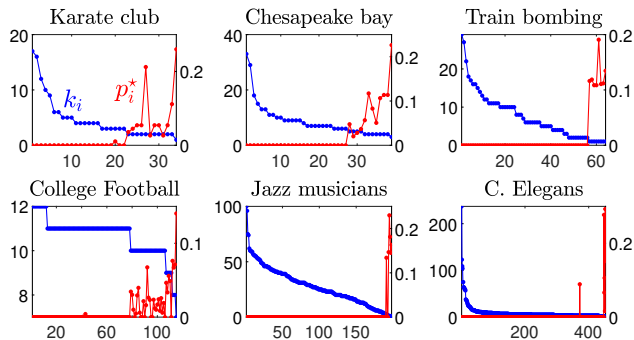}
    \caption{Degree sequence (blue) and maximum variance distribution (red) on a number of real-world networks from the KONECT database \cite{Kunegis_KONECT}. The horizontal axis contains node indices $i$, sorted according to decreasing degree. For each node, the degree $k_i$ and maximum variance probability $p^\star_i$ are given, with scales on the left and right vertical axis, respectively.}
    \label{fig: degrees and maxvar distributions}
\end{figure}

\section{Alternative formulations of locally optimal solutions}\label{S: alternative formulations}
We present a number of equivalent formulations for the local optimality condition \eqref{eq: 2 local optimality condition}. These reformulations are relevant in the context of our article since they determine the form of the maximum variance solution $\mathbf{p}^\star$, but more importantly, the vector $\mathbf{p}$ that solves \eqref{eq: 2 local optimality condition} (with $\alpha=2\var_{\omega}(\mathbf{p})$) uniquely for a given graph seems to be a natural algebraic object on this graph, worthy of a more thorough characterization. The many different formulations in Proposition \ref{propos: equivalent characterizations} below are an attest to the different settings in which the vector $\mathbf{p}$ seems to play a role. Furthermore, as we discuss at the end of this section, equation \eqref{eq: 2 local optimality condition} and its solution are related to the theory of magnitude \cite{Leinster_book}.
\\
We first introduce two results of Miroslav Fiedler from which a distinct \emph{geometric} and \emph{algebraic} characteristic of $\mathbf{p}$ will follow. In his book \cite{Fiedler_matrices_and_graphs}, Fiedler collects a broad range of results that follow from a geometric perspective on graphs. At the core of this perspective is the fact that each graph has an embedding $\mathbf{m}:\mathcal{N}\rightarrow \mathbb{R}^{n-1}$ into Euclidean space such that $$
\Vert\mathbf{m}(i)-\mathbf{m}(j)\Vert^2 = \omega_{ij}\text{~for all $i,j\in\mathcal{N}$},
$$
where the embedded nodes are the vertices of a hyperacute simplex (see \cite{Fiedler_matrices_and_graphs, krl_simplex} for more details). In other words $\mathbf{m}$ is an isometric embedding of the node set with respect to the square root resistance distance. This embedding can be extended to functions on the nodes as $\mathbf{m}(\mathbf{q})=\sum_{i\in\mathcal{N}}q_i\mathbf{m}(i)
$ which means that we can interpret node functions as `coordinates' of points in $\mathbb{R}^{n-1}$. From this perspective, we will see that $\mathbf{p}$ is the coordinate of a distinguished point of the simplex with vertices $\mathbf{m}(\mathcal{N})$.
\\
A second place where equation \eqref{eq: 2 local optimality condition} appears in the work of Fiedler is in a matrix identity between the effective resistance and Laplacian matrices. Writing the local optimality condition \eqref{eq: 2 local optimality condition} in block-matrix form, we find \small{$\begin{pmatrix}0&\mathbf{u}^T\\\mathbf{u}&\Omega\end{pmatrix}\begin{pmatrix}-2\var_{\omega}(\mathbf{p})\\\mathbf{p}\end{pmatrix}=\begin{pmatrix}1\\\mathbf{0}\end{pmatrix}$}\normalsize  with zero vector $\mathbf{0}=(0,\dots,0)^T$. The following result of Fiedler shows that this equation can be completed to a full (inverse) matrix identity:
\begin{theorem}[Fiedler's identity]
The Laplacian matrix $Q$ and resistance matrix $\Omega$ of a graph satisfy the matrix identity
\begin{equation}\label{eq: Fiedlers identity}
\begin{pmatrix}
0&\mathbf{u}^T\\\mathbf{u}&\Omega
\end{pmatrix}^{-1} = -\frac{1}{2}\begin{pmatrix}
4\var_{\omega}(\mathbf{p})&-2\mathbf{p}^T\\-2\mathbf{p}&Q
\end{pmatrix}
\end{equation}
where $\Omega\mathbf{p}=2\var_{\omega}(\mathbf{p})\mathbf{u}$.
\end{theorem}
\textbf{Proof:} See proof in \cite[Thm. 1.4.1]{Fiedler_matrices_and_graphs}. The validity of equation \eqref{eq: Fiedlers identity} can also be checked by multiplying both matrices, which results in the identity matrix. \hfill$\square$
\\
Using the embedding $\mathbf{m}$ of a graph and Fiedler's identity \eqref{eq: Fiedlers identity} we now find the following reformulations of equation \eqref{eq: 2 local optimality condition}:
\begin{proposition}\label{propos: equivalent characterizations}
The following are equivalent characterizations for a vector $\mathbf{p}\in\mathbb{R}^n$ and its variance $\var_{\omega}(\mathbf{p})$:
\begin{itemize}
    \item[(i)] \textup{$\Omega\mathbf{p} = 2\var_{\omega}(\mathbf{p})\mathbf{u}$. In other words, $\mathbf{p}$ solves equation \eqref{eq: 2 local optimality condition} with $\alpha=2\var_{\omega}(\mathbf{p})$}.
    \item[(ii)] $\mathbf{p}=\operatorname{argmax}_{\mathbf{q}}\big\lbrace\frac{1}{2}\mathbf{q}^T\Omega\mathbf{q}$\textup{~s.t.~}$\mathbf{u}^T\mathbf{q}=1\big\rbrace$\textup{, with corresponding optimal value $\var_{\omega}(\mathbf{p})$}
    \item[(iii)] \textup{$\Vert\mathbf{m}(\mathbf{p})-\mathbf{m}(i)\Vert^2 = \var_{\omega}(\mathbf{p})$ for all $i\in\mathcal{N}$ and isometric embedding $\mathbf{m}$. In other words, $\mathbf{m}(\mathbf{p})$ is the \emph{circumcenter} of the sphere going through vertices $\mathbf{m}(\mathcal{N})$, with \emph{circumradius} $\sqrt{\var_{\omega}(\mathbf{p})}$.}
    \item[(iv)] \textup{$\mathbf{p}=\frac{1}{2}Q\zeta + \mathbf{u}/n$, with corresponding variance $\var_{\omega}(\mathbf{p})=\tfrac{1}{4}\zeta^T Q\zeta+\mathbf{u}^T\zeta/n$, where $\zeta=\diag(Q^\dagger)$}
    \item[(v)] \textup{$p_i = 1-\frac{1}{2}\sum\limits_{j\sim i}c_{ij}\omega_{ij}$ for all $i\in\mathcal{N}$}
\end{itemize}
\end{proposition}
\textbf{Proof:} The proof is given in Appendix \ref{A: proof of reformulations}. We remark that the characterizations (i) and (iii) are discussed in \cite{Fiedler_matrices_and_graphs}, and the form (v) appears in \cite{Bapat_inverse_of_weighted_graphs}. In formulation (v), the normalization of $p_i$ is guaranteed by $\sum_{j\sim i}c_{ij}\omega_{ij}=n-1$, which is known as Foster's Theorem \cite{Foster}.\hfill$\square$
\\
While formulated in terms of solutions of the local optimality condition on the full node set $\mathcal{N}$, Proposition \ref{propos: equivalent characterizations} above in fact describes the solutions for general $\mathcal{V}$ due to an important recursive structure of the effective resistance: for any resistance matrix $\Omega$ of a graph $G$, the submatrix $\Omega_{\mathcal{V}\mathcal{V}}$ is again the resistance matrix of a graph $G'$. More precisely, if we write the Laplacian matrix in block-form $Q=\left(\begin{smallmatrix}Q_{\mathcal{V}\mathcal{V}}&Q_{\mathcal{V}\mathcal{V}^c}\\Q_{\mathcal{V}^c\mathcal{V}}&Q_{\mathcal{V}^c\mathcal{V}^c}\end{smallmatrix}\right)$ for some node-set $\mathcal{V}$ and its complement $\mathcal{V}^c$, then the \emph{Schur complement of $Q$ with respect to $\mathcal{V}$} is defined as
\begin{equation}\label{eq: Schur complement}
Q/\mathcal{V}^c \triangleq Q_{\mathcal{V}\mathcal{V}} - Q_{\mathcal{V}\mathcal{V}^c}(Q_{\mathcal{V}^c\mathcal{V}^c})^{-1}Q_{\mathcal{V}^c\mathcal{V}}
\end{equation}
and has the important properties that it is again a Laplacian matrix of some graph $G'$ and that this graph has resistance matrix $\Omega'=\Omega_{\mathcal{V}\mathcal{V}}$, see \cite{Fiedler_matrices_and_graphs, Dorfler_Kron_reduction}. Consequently, the solutions described in Proposition \ref{propos: equivalent characterizations} are valid for the local optimum on any set $\mathcal{V}$ by considering the Laplacian $Q/\mathcal{V}^c$ and effective resistance matrix $\Omega_{\mathcal{V}\mathcal{V}}$ for which $\mathcal{V}$ is the full node set. Due to this recursive structure of Laplacians and the effective resistance, we can intuitively understand the method of solving the maximum variance problem as calculating the variance of the locally optimal solution $\mathbf{p}_{\mathcal{V}}$ for each set $\mathcal{V}$ by \eqref{eq: 2 local optimality condition} and then choosing the feasible ones (i.e. distributions) by \eqref{eq: 1 nonnegative optimality condition} and characterizing the globally optimal one by condition \eqref{eq: 3 global optimality condition}. In this perspective, we have a last result that relates solutions of \eqref{eq: 2 local optimality condition} and their variances between different sets of nodes:
\begin{proposition}\label{propos: relation between maxvar distributions}
The solution $\mathbf{p}$ to equation \eqref{eq: 2 local optimality condition} for the set $\mathcal{V}=\mathcal{N}$ and $\mathbf{p}'$ for the sets $\mathcal{V}=\mathcal{N}\backslash\lbrace x\rbrace$ are related by
$$
\begin{cases}
\mathbf{p}' = \mathbf{p} - \frac{p_x}{k_x}\sum_{j\sim x}c_{xj}(e_x-e_j)\\
\var_{\omega}(\mathbf{p}')=\var_{\omega}(\mathbf{p})-\frac{p^2_x}{k_x}
\end{cases}
$$
\end{proposition}
\textbf{Proof:} Proposition \ref{propos: relation between maxvar distributions} is proven in Appendix \ref{A: proof of reformulations} making use of the Schur complement of the graph Laplacian and Fiedler's identity.\hfill$\square$
\\
In particular, Proposition \ref{propos: relation between maxvar distributions} thus shows that if a locally optimal distribution is non-negative for a set $\mathcal{V}$ as $\mathbf{p}_{\mathcal{V}}\geq 0$, then the locally optimal solutions for all subsets $\mathcal{W}\subset\mathcal{V}$ will be non-negative as well, by positivity of the factor $p_xc_{xj}/k_x$. Furthermore, we find that the variance $\var_{\omega}$ of locally optimal distributions is non-increasing when considering subsets. 
\\~\\
The \emph{magnitude} as introduced by Leinster \cite{Leinster_book} is an invariant that can be defined for various mathematical objects such as enriched categories, metric spaces or graphs, and captures a notion of \emph{size} of the respective objects. Most simply, let $\mathcal{X}$ be a set with a notion of similarity $z:\mathcal{X}\times\mathcal{X}\rightarrow\mathbb{R}$ between its elements, which can be captured in a matrix $(Z)_{ij}=z(i,j)$. The \emph{weight vector} $\mathbf{q}$ of this space if then calculated from $Z\mathbf{q}=\mathbf{u}$, if this solution exists, and the \emph{magnitude} is then defined by $\vert Z\vert = \sum_{i\in\mathcal{X}} q_i$. In the case where $\mathcal{X}$ is a metric space, one may take $z(i,j)=1-d(i,j)$ (other mappings exist) and the magnitude is then expressed in terms of the distance matrix. For graphs equipped with the resistance distance $(\mathcal{N},\omega)$ this mapping gives the correspondence $\vert Z\vert = (1-2\var_{\omega}(\mathbf{p}))^{-1}$ and $\mathbf{q}=(1-2\var_{\omega}(\mathbf{p}))^{-1}\mathbf{p}$ between magnitude and weight vector on one side, and vector $\mathbf{p}$ and its variance on the other. As the theory of magnitude is well-studied, this relation could be an interesting starting point for further research into the properties of $\mathbf{p}$ and its variance.


\section{Variance of diffusion processes and an alternative distance measure}\label{A: variance for diffusion}
A well-studied dynamical process on networks is the diffusion process, where a time-dependent state $\mathbf{p}_t$ is defined on the nodes of a graph with Laplacian $Q$ and evolves according to the diffusion equation (also called heat equation)
$$
\frac{d}{dt}\mathbf{p}_t = -Q\mathbf{p}_t\text{~with solution~}\mathbf{p}_t = e^{-Qt}\mathbf{p}_0\text{~for $t>0$}
$$
for some initial state $\mathbf{p}_0$. From properties of the Laplacian matrix it follows that if the initial state of the process is a distribution then all further states will be distributions as well, i.e. $\mathbf{p}_0\in\Delta_n\Rightarrow \mathbf{p}_{t}\in\Delta_n$ for all $t>0$, with the uniform distribution $p_{\infty}=\mathbf{u}/n$ as eventual stationary state.
\\
We now consider the variance $\var_{\omega}(\mathbf{p}_t)$ of the time-evolving distribution with respect to the square root effective resistance \eqref{eq: definition resistance variance and covariance}. Using the fact that $\Omega Q = 2I-2\mathbf{u}\mathbf{p}^T$ (e.g. following \eqref{eq: Fiedlers identity}) with $\mathbf{p}$ as in Proposition \ref{propos: equivalent characterizations}, we find that this variance evolves as
$$
\frac{d}{dt}\var_{\omega}(\mathbf{p}_t) = 2\Vert \mathbf{p}_t\Vert_2^2 - 2\mathbf{p}^T\mathbf{p}_t.
$$
In this expression, the first term causes the variance to increase, while the second term causes a decreasing variance depending on how similar (in the inner-product sense) the probability at time $t$ is to the vector $\mathbf{p}$. This effectively separates the probability simplex into two regions
\begin{align*}
&\Delta_n^+ = \lbrace \mathbf{q}\in\Delta_n:\Vert\mathbf{q}\Vert^2>\mathbf{p}^T\mathbf{q}\rbrace\text{, and~}
\\
&\Delta_n^- = \lbrace \mathbf{q}\in\Delta_n:\Vert\mathbf{q}\Vert^2<\mathbf{p}^T\mathbf{q}\rbrace
\end{align*}
where the variance either increases or decreases for the diffusion process. 
\\
In the particular case of node-transitive graphs, we find a simplified evolution of the variance
$d\var_{\omega}(\mathbf{p}_t)/dt = 2\Vert\mathbf{p}_t\Vert^2-2/n$ due to the fact that $\mathbf{p}=\mathbf{u}/n$ (see proof of Proposition \ref{prop: maxvar on node transitive}). Since $\Vert\mathbf{q}\Vert^2\geq 1/n$ for all $\mathbf{q}\in\Delta_n$ with equality only for the uniform distribution, this shows that the diffusion process results in a strictly increasing variance up to the stationary state. In the case of arbitrary graphs, a similar evolution is reproduced if we consider the variance with respect to an alternative dissimilarity $d$ in definition \eqref{eq: definition graph variance} as $\var_{Q^\dagger}(\mathbf{p}) = -\mathbf{p}^TQ^\dagger\mathbf{p}$. The time-evolution in this case is given by
$$
\frac{d}{dt}\var_{Q^\dagger}(\mathbf{p}_t) = 2\Vert\mathbf{p}_t\Vert_2^2 - 2/n \geq 0
$$
with equality if and only if $\mathbf{p}_t=\mathbf{p}_{\infty}=\mathbf{u}/n$. In other words, the diffusion process results in a time-increasing variance $\var_{Q^\dagger}$ up to the uniform stationary distribution, when the variance is maximised. Note the important differences between the two choices $\var_{\omega}$ and $\var_{Q^\dagger}$, whose maximum distribution accumulates probability at the extremities of the graph in the former case, and spreads it uniformly in the latter case.

\section{Maximum variance support for tree graphs}\label{A: maxvar for tree graphs}
We show Proposition \ref{propos: maxvar support on trees} which says that the maximum variance distribution on tree graphs is supported on the leaf nodes. This result was mentioned in \cite{Hjorth_finite_metric_spaces} without proof, and was proven in \cite{Dankelman_average_distance} based on a perturbation argument of the maximum variance distribution for an equivalent problem formulation.
\\
Here, we present a proof by showing that having a non-leaf node in the maximum variance support contradicts the necessary conditions \eqref{eq: 1 nonnegative optimality condition}$-$\eqref{eq: 3 global optimality condition} for the maximum variance distribution. For our proof, we will make use of the fact that the effective resistance on trees corresponds to the shortest path distance (see e.g. \cite{Klein_resistance_distance}) and thus that
\begin{equation}\label{eq: triangle equality for hinges}
\omega_{ij}=\omega_{ix}+\omega_{xj}
\end{equation}
whenever the (unique) path from $i$ to $j$ passes through node $x$ or, equivalently, when removing node $x$ from the graph disconnects $i$ and $j$. We will consider an arbitrary non-leaf node $x$ in the tree graph and one of its neighbours $y\sim x$. We write $\mathcal{N}_y$ for the connected set of nodes after removal of $x$ that contains node $y$, and write $\mathcal{V}_y=\mathcal{N}_y\cap\mathcal{V}^\star$ (which could be empty) and $\mathcal{V}_y^c=\mathcal{V}^\star\backslash\mathcal{V}_y$ (which always contains $x$) as in the figure below. Following \eqref{eq: triangle equality for hinges}, we thus have $\omega_{iy}=\omega_{ix}+\omega_{xy}$ for all $i\in\mathcal{V}_y^c$ and since removing $y$ disconnects $\mathcal{V}_y$ from $x$ we also have $\omega_{iy}=\omega_{ix}-\omega_{xy}$ for all $i\in\mathcal{V}_y$. 
\begin{figure}[h!]
    \centering
    \includegraphics{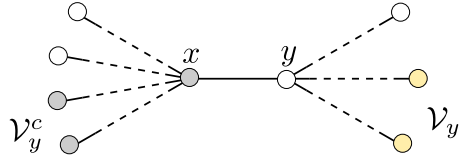}
    \caption{Node sets $\mathcal{V}_x,\mathcal{V}_y$ corresponding to a link $x\sim y$.}
\end{figure}
We now give the proof by contradiction.
\\
\textbf{Proof of Proposition \ref{propos: maxvar support on trees}:} Assume that a non-leaf node $x$ has non-zero probability in the maximum variance distribution $\mathbf{p}^\star$, i.e. $x\in\mathcal{V}^\star$, and let $y$ be a neighbour of $x$.
\\
We first show that $\mathcal{V}_y$ is non-empty. Assume otherwise (in particular $y\notin\mathcal{V}^\star$), then we can write
\begin{align*}
&\sum_{i\in\mathcal{V}^\star(=\mathcal{V}^c_y)}p^\star_i\omega_{iy} - \sum_{k\in\mathcal{V}^\star}p^\star_k\sum_{i\in\mathcal{V}^\star}p^\star_i\omega_{ik}
\\
&=\sum_{i\in\mathcal{V}^c_y}p^\star_i(\omega_{ix}+\omega_{xy}) - \sum_{k\in\mathcal{V}^\star}p^\star_k\sum_{i\in\mathcal{V}^\star}p^\star_i\omega_{ik}
\\
&=\omega_{xy} - \sum_{k\in\mathcal{V}^\star}p^\star_k\sum_{i\in\mathcal{V}^\star}p^\star_i(\omega_{ik}-\omega_{ix})
\\
&=\omega_{xy}
\end{align*}
where the last step uses \eqref{eq: 2 local optimality condition} as $\sum_{i\in\mathcal{V}^\star}p^\star_i\omega_{ik} = \sum_{i\in\mathcal{V}^\star}p^\star_i\omega_{ix}$. Invoking the global optimality condition we then find \eqref{eq: 3 global optimality condition}$\Leftrightarrow \omega_{xy}< 0$ which contradicts the fact that $\omega_{xy}>0$. Consequently, for every neighbour $y$ of $x$ the set $\mathcal{V}_y$ must be non-empty.
\\~\\
Next, we consider two cases for the neighbour $y$. (Case 1) If $y\notin\mathcal{V}^\star$ then we can write
\begin{align*}
&\sum_{i\in\mathcal{V}^\star}p^\star_i\omega_{iy} - \sum_{k\in\mathcal{V}^\star}p^\star_k\sum_{i\in\mathcal{V}^\star}p^\star_i\omega_{ik}
\\
&=\sum_{i\in\mathcal{V}_y}p^\star_i(\omega_{ix}-\omega_{xy}) + \sum_{i\in\mathcal{V}^c_y}p^\star_i(\omega_{ix}+\omega_{xy}) 
\\
&\hphantom{=}- \sum_{k\in\mathcal{V}^\star}p^\star_k\sum_{i\in\mathcal{V}^\star}p^\star_i\omega_{ik}
\\
&=\omega_{xy}\left(\sum_{i\in\mathcal{V}^c_y}p^\star_i - \sum_{i\in\mathcal{V}_y}p^\star_i\right)
\\
&=\omega_{xy}\left(1-2\sum_{i\in\mathcal{V}_y}p^\star_i\right)
\end{align*}
The global optimality condition then gives that
$$
\text{\eqref{eq: 3 global optimality condition}}\Leftrightarrow \sum_{i\in\mathcal{V}_y}p^\star_i> \frac{1}{2}\text{~for all $y\sim x$ with $y\notin\mathcal{V}^\star$}
$$
(Case 2) If $y\in\mathcal{V}^\star$ then we can write
\begin{align*}
    &\sum_{i\in\mathcal{V}^\star}p^\star_i\omega_{ix} - \sum_{i\in\mathcal{V}^\star}p^\star_i\omega_{iy}
    \\
    &=\sum_{i\in\mathcal{V}_y}p^\star_i(\omega_{iy}+\omega_{ix}) + \sum_{i\in\mathcal{V}_y^c}p^\star_i(\omega_{iy}-\omega_{ix}) - \sum_{i\in\mathcal{V}^\star}p^\star_i\omega_{iy}
    \\
    &=\sum_{i\in\mathcal{V}^c_y}p^\star_i - \sum_{i\in\mathcal{V}_y}p^\star_i 
    \\
    &=1-2\sum_{i\in\mathcal{V}^c_y}p^\star_i
\end{align*}
The local optimality condition then gives
$$
\text{\eqref{eq: 2 local optimality condition}}\Leftrightarrow\sum_{i\in\mathcal{V}_y}p^\star_i = \frac{1}{2}\text{~for all $y\sim x$ with $y\in\mathcal{V}^\star$}
$$
Since each neighbour is either in the maximum variance support (Case 2) or not (Case 1), we have that
$$
1=\sum_{i\in\mathcal{V}^\star}p^\star_i = p^\star_x+\sum_{y\sim x}\sum_{i\in\mathcal{V}_y}p^\star_i>\sum_{y\sim x}\frac{1}{2}
$$
which shows that $x$ can have at most one neighbour. This contradicts our initial assumption that $x$ is a non-leaf node and thus shows that no maximum variance distribution (satisfying \eqref{eq: 1 nonnegative optimality condition}$-$\eqref{eq: 3 global optimality condition}) can have a non-leaf node in the support, hence $\mathcal{V}^\star$ must be a subset of the leaf nodes.\hfill$\square$

\section{Maximum variance distribution on weighted stars and configuration graphs}\label{A: maxvar for stars and config graphs}
We show Proposition \ref{prop: maximum variance support weighted star} and Corollary \ref{cor: maximum variance support configuration graph} that characterize the maximum variance distributions on weighted star graphs and configuration graphs, respectively. We first derive the solution for the weighted star using properties of tree graphs, and then show an equivalence to the maximum variance problem on configuration graphs which leads to the corollary.
\\
We start with the following result about the solution to equation \eqref{eq: 2 local optimality condition} for (the leaf nodes of) a weighted star:
\begin{lemma}\label{lem: p for weighted star}
For any subset $\mathcal{V}$ of the leaf nodes of a weighted star, the solution $\mathbf{p}$ to equation \eqref{eq: 2 local optimality condition} is given by
\begin{equation}\label{eq: p for weighted star}
p_i = \frac{1}{2} - \frac{(v-2)k_i}{4m_\mathcal{V}}\text{~for all $i\in\mathcal{V}$}
\end{equation}
where $2m_{\mathcal{V}}=\sum_{i\in\mathcal{V}}k_i$.
\end{lemma}
\textbf{Proof:} We can assume without loss of generality that $\mathcal{V}$ is the full set of leaf nodes. If $\mathcal{V}$ is a subset of the leaf nodes, we can remove those leaf nodes not in $\mathcal{V}$ and obtain a weighted star (parametrized by degrees $\lbrace k_i\rbrace_{i\in\mathcal{V}})$ with leaf nodes $\mathcal{V}$, to which the Lemma is then applicable.
\\
As pointed out in Appendix \ref{A: maxvar for tree graphs}, the effective resistance on tree graphs (and thus weighted stars) equals the shortest path distance. For the leaf nodes in a weighted star, which are connected by a single, weighted link to node $0$, we thus find that $\omega_{io}=c_{io}^{-1}=k_i^{-1}$. Furthermore, for pairs of leaf nodes we find
$$
\omega_{ij} =\omega_{i0}+\omega_{0j}= k_{i}^{-1} + k_j^{-1}\text{~for leaf nodes $i,j$}
$$
since removing $\lbrace 0\rbrace$ disconnects all leaf nodes. If we let $\tilde{\mathbf{k}}=(k_1^{-1},\dots,k_n^{-1})$ we can thus write the resistance matrix between leaf nodes $\mathcal{V}$ as 
$$
\Omega_{\mathcal{V}\mathcal{V}} = \tilde{\mathbf{k}}\mathbf{u}^T + \mathbf{u}\tilde{\mathbf{k}}^T - 2\operatorname{diag}(\tilde{\mathbf{k}}).
$$
We now show that the local optimality condition $\Omega_{\mathcal{V}\mathcal{V}}\mathbf{p}=(2\sigma^2)\mathbf{u}$ for the maximum variance solution $\mathbf{p}$ is satisfied by the proposed solution. Introducing solution \eqref{eq: p for weighted star} in vector form into condition \eqref{eq: 2 local optimality condition}, we find
\begin{align*}
\Omega_{\mathcal{V}\mathcal{V}}\mathbf{p} &= (   \tilde{\mathbf{k}}\mathbf{u}^T + \mathbf{u}\tilde{\mathbf{k}}^T - 2\operatorname{diag}(\tilde{\mathbf{k}}))\left[\frac{1}{2}\mathbf{u} - \frac{(v-2)}{4m_{\mathcal{V}}}\mathbf{k}\right]
\\
&=\left[\frac{v}{2} - \frac{(v-2)}{2} - 1\right]\tilde{\mathbf{k}} 
\\
&\hphantom{=}+\left[\frac{\tilde{\mathbf{k}}^T\mathbf{u}}{2} - \frac{v(v-2)}{4m_{\mathcal{V}}} + \frac{(v-2)}{2m_{\mathcal{V}}}\right]\mathbf{u}
\\
&= \left[\frac{\tilde{\mathbf{k}}^T\mathbf{u}}{2} - \frac{(v-2)^2}{2m_{\mathcal{V}}}\right]\mathbf{u}
\end{align*}
as required, which completes the proof. Alternatively, \eqref{eq: p for weighted star} can be derived from equation $\Omega_{\mathcal{V}\mathcal{V}}\mathbf{p}=2\sigma^2\mathbf{u}$ by introducing the found effective resistance matrix. \hfill$\square$ 
\\
We can now continue to prove Proposition \ref{prop: maximum variance support weighted star}, by showing that the solution $\mathbf{p}$ to equation \eqref{eq: 2 local optimality condition} described in Lemma \ref{lem: p for weighted star} together with conditions \eqref{eq: 1 nonnegative optimality condition} and \eqref{eq: 3 global optimality condition} are equivalent to the conditions described in the proposition. We assume all degrees to be ordered in \emph{increasing} order, i.e. with $k_i\leq k_{i+1}$ for all $i<n$.
\\
\textbf{Proof of Proposition \ref{prop: maximum variance support weighted star}:} From Lemma \ref{lem: p for weighted star} we know that the local optimality condition \eqref{eq: 2 local optimality condition} is equivalent to \eqref{eq: p for weighted star}. We now consider when the two other necessary and sufficient conditions are met.
\\
Introducing expression \eqref{eq: p for weighted star} for the solution to the local optimality condition into the positivity condition \eqref{eq: 1 nonnegative optimality condition} for $\mathbf{p}$, we find that
$$
\text{\eqref{eq: 1 nonnegative optimality condition}}\Leftrightarrow k_i \leq \frac{\sum_{j\in\mathcal{V}} k_j}{v-2}\text{~for all $i\in\mathcal{V}$}.
$$
Next, if $\mathcal{V}$ is a subset of the leaf nodes we let $j\notin\mathcal{V}$ be one of the leaf nodes not in the support, and write
\begin{align*}
    &\sum_{i\in\mathcal{V}}p_i\omega_{ij} - \sum_{i,r\in\mathcal{V}}p_ip_r\omega_{ir}
    \\
    &=\sum_{i\in\mathcal{V}}p_i(k_i^{-1}+k_j^{-1}) - \sum_{i,r\in\mathcal{V}}p_ip_r(k_i^{-1}+k_r^{-1}) + 2\sum_{i\in\mathcal{V}}p_i^2k_i^{-1}
    \\
    &=k_j^{-1} - \sum_{i\in\mathcal{V}}p_ik_i^{-1} + 2\sum_{i\in\mathcal{V}}p_i^2k_i^{-1}
    \\
    &= k_j^{-1} -2 \sum_{i\in\mathcal{V}}p_ik_i^{-1}\left(\frac{1}{2}-p_i\right)
    \\
    &= k_j^{-1} - \frac{(v-2)}{2m_{\mathcal{V}}}\sum_{i\in\mathcal{V}}p_i = k_j^{-1} - \frac{(v-2)}{2m_{\mathcal{V}}}.
\end{align*}
Consequently, the global optimality condition \eqref{eq: 3 global optimality condition} is equivalent to
$$
\text{\eqref{eq: 3 global optimality condition}}\Leftrightarrow \frac{\sum_{j\in\mathcal{V}}k_j}{v-2}< k_i\text{~for all~$i\notin\mathcal{V}$}
$$
Taking into account the most extreme cases, the optimality conditions \eqref{eq: 1 nonnegative optimality condition}$-$\eqref{eq: 3 global optimality condition} for the maximum variance set $\mathcal{V}^\star$ are thus
$$
\max_{i\in\mathcal{V}^\star} k_i \leq \frac{\sum_{j\in\mathcal{V}^\star}k_j}{v-2} < \min_{i\notin\mathcal{V}^\star} k_i
$$
which in particular shows that $k_i< k_j$ for all $i\in\mathcal{V}^\star$ and $j\notin\mathcal{V}^\star$ and thus that $\mathcal{V}^\star$ can only be equal to a set of nodes with the $\ell$ smallest degrees. More precisely, making use of the ordering we find that $\mathcal{V}^\star = \lbrace i\in\mathcal{N}: k_i\leq k_\ell\rbrace$ where $k_\ell$ satisfies
\begin{equation}\label{eq: equation for ell}
k_\ell\leq \frac{\sum_{i=1}^\ell k_i}{\ell-2}< k_{\ell+1}
\end{equation}
and defining $k_{n+1}=\infty$ for consistency. The condition \eqref{eq: equation for ell} for $\ell$ can be satisfied by only one value of at the same time. For contradiction, assume there are two such values $\ell<m$, then we would have
$$
k_{m} \overset{(a)}{\leq} \frac{\sum_{i=1}^m k_i}{m-2} \overset{(b)}{\leq} \frac{\sum_{i=1}^\ell k_i}{m-2} + \frac{(m-\ell)}{m-2}k_m \overset{(c)}{<}\left(\frac{\ell-2}{m-2}+\frac{m-\ell}{m-2}\right)k_m=k_m
$$
where (a) is the lower bound from equation \eqref{eq: equation for ell} with index $m$, (b) splits up the sum in the numerator and uses that $k_i\leq k_m$ for all $\ell< i\leq m$ and (c) is the upper bound from equation \eqref{eq: equation for ell} with index $\ell$; since this produces a contradiction we know that $\ell$ must be unique. Existence of a solution to equation \eqref{eq: equation for ell} for at least one index follows from the observation that the lower bound in \eqref{eq: equation for ell} holds for $\ell=3$ combined with 
$$
\frac{\sum_{i=1}^\ell k_i}{\ell-2} \geq k_{\ell+1} \Rightarrow \frac{\sum_{i=1}^{\ell+1}k_i}{\ell-1}\geq k_{\ell-1}.
$$
In other words if the upper bound in \eqref{eq: equation for ell} is violated for some $\ell$, then the lower bound is satisfied for $\ell+1$. We then inductively find that the lower bound is always satisfied for $\ell\geq 3$, either until the upper bound is also satisfied for some $\ell<n$ and we thus have found a solution to \eqref{eq: equation for ell}, or until $\ell=n$ for which the upper bound is always satisfied. Thus there exists a unique solution to \eqref{eq: equation for ell} and since the maximum variance support is a subset of the leaf nodes (by Proposition \ref{propos: maxvar support on trees}), this fully characterizes the maximum variance support $\mathcal{V}^\star$. \hfill$\square$
\\
We now show that the result for weighted star graphs immediately implies the result for configuration graphs, based on the following correspondence between effective resistances in both graphs
\begin{lemma}
The effective resistance $\omega_{ij}$ between pairs of nodes in a configuration graph and $\omega'_{ij}$ between pairs of leaf nodes in a weighted star parametrized by the same degree sequence $\mathbf{k}$ is proportional:
\begin{equation}
\omega_{ij} = \frac{2m-1}{2m} \omega'_{ij}
\end{equation}
where $i$ and $j$ correspond to nodes with the same degrees in both graphs.
\end{lemma}
\textbf{Proof:} Our proof will make use of the Schur complement. An important property of the Schur complement of a Laplacian matrix is that it leaves the effective resistances invariant \cite{Dorfler_Kron_reduction, karel_resistance>distance}, see also Section \ref{S: alternative formulations}. More precisely, if $\Omega$ is the effective resistance matrix of a Laplacian matrix $Q$, then the Schur complement $Q/\mathcal{V}^c$ with respect to some set $\mathcal{V}$ has effective resistance matrix $\Omega_{\mathcal{V}\mathcal{V}}$ equal to the submatrix of $\Omega$ with rows and columns in $\mathcal{V}$.
\\
The Laplacian $Q'$ of the weighted star can be written in block-form as
$$
Q' = \begin{pmatrix}2m & -\mathbf{k}^T\\-\mathbf{k}&\diag(\mathbf{k})
\end{pmatrix}
$$
where the first row and column correspond to node $\lbrace 0\rbrace$ in the star. Taking the Schur complement with respect to the leaf nodes $\lbrace 0\rbrace^c$ we find 
$$
Q'/\lbrace 0\rbrace = \diag(\mathbf{k}) - \mathbf{k}\mathbf{k}^T/2m = \frac{2m-1}{2m}Q_{\text{conf.}}
$$
with $Q_{\text{conf.}}$ the Laplacian of the configuration graph. Since the Schur complement leaves the effective resistances invariant, we thus find that 
$$
\Omega'_{\mathcal{V}\mathcal{V}} = \frac{2m}{2m-1}\Omega_{\text{conf.}}
$$
as required. \hfill$\square$
\\
Corollary \ref{cor: maximum variance support configuration graph} now follows as a simply corollary of Lemma \ref{lem: p for weighted star} and Proposition \ref{prop: maximum variance support weighted star} combined:
\\
\textbf{Proof of Corollary \ref{cor: maximum variance support configuration graph}:} Since the maximum variance distribution on a weighted star is always supported on the leaf nodes, the maximum variance problem \eqref{eq: maximum variance problem} for the distance matrix $\Omega'$ of the weighted star is the same as the maximum variance problem on the distance matrix $\Omega'_{\mathcal{V}\mathcal{V}}$, which is equivalent to the optimization problem on the rescaled distance matrix $\Omega$ of the configuration graph, up to a scaling of $\frac{2m-1}{2m}$ of the maximum variance solution $2\sigma^2$. The criterion for the maximum variance support $\mathcal{V}^\star$ for weighted star graphs given in Proposition \ref{prop: maximum variance support weighted star} thus also describes the maximum variance support for configuration graphs as in Corollary \ref{cor: maximum variance support configuration graph}.\hfill$\square$

\section{Proof of Proposition \ref{propos: equivalent characterizations} on solutions to equation \eqref{eq: 2 local optimality condition}}\label{A: proof of reformulations} We prove Proposition \ref{propos: equivalent characterizations} which describes equivalent characterizations for a vector $\mathbf{p}\in\mathbf{R}^n$.
\\
\textbf{Proof of Proposition \ref{propos: equivalent characterizations}:} \textbf{(i)$\Leftrightarrow$(ii)} Using a similar derivation as in Section \ref{A: proof of necessary and sufficient conditions} which considers a transfer of probability $\epsilon$ between nodes $i$ and $j$, we find that a necessary condition for any solution $\mathbf{p}$ to the optimization problem in (ii) is that $\Omega\mathbf{p}=2\var_{\omega}(\mathbf{p})\mathbf{u}$, in other words (ii)$\Rightarrow$(i). Since there is a unique solution to (i) this means it is also necessary and (i)$\Rightarrow$(ii). 
\\
\textbf{(i)$\Leftrightarrow$(iii)} To prove this result, we make use of a particular embedding $\mathbf{m}$ based on the $(n-1)$ eigenvectors $\mathbf{z}_k\in\mathbb{R}^n$ with $\mathbf{z}_k\perp\mathbf{u}$ and non-zero eigenvalues $\mu_k$ of the Laplacian $Q$ as (see \cite{krl_simplex})
$$
(\mathbf{m}(i))_k = (\mathbf{z}_k)_i\mu_k^{-1/2}\text{~for all $i\in\mathcal{N}$ and $k<n$}.
$$
This embedding satisfies $\mathbf{m}(i)^T\mathbf{m}(j) = (Q^\dagger)_{ij}$ for all $i,j\in\mathcal{N}$, which means we can write the distance from a point $\mathbf{m}(\mathbf{p})$ to a vertex as
$$
\Vert \mathbf{m}(\mathbf{p})-\mathbf{m}(i)\Vert^2 = (e_i-\mathbf{p})^TQ^\dagger(e_i-\mathbf{p})
$$
We then find that
\begin{align*}
&\Vert \mathbf{m}(\mathbf{p})-\mathbf{m}(i)\Vert^2 - \Vert \mathbf{m}(\mathbf{p})-\mathbf{m}(j)\Vert^2
\\
&= e_i^TQ^\dagger e_i - e_j^TQ^\dagger e_j -2\mathbf{p}^T Q^\dagger (e_i-e_j)
\\
&= (e_i-e_j)^T\Omega\mathbf{p}
\end{align*}
which shows that the distance from $\mathbf{m}(\mathbf{p})$ is the same to all vertices $\mathbf{m}(i)$ if and only if $\Omega\mathbf{p}$ is a constant vector. In this case, the radius of the circumsphere can furthermore be calculated as
\begin{align*}
\Vert\mathbf{m}(\mathbf{p})-\mathbf{m}(i)\Vert^2 &=
(\mathbf{p}-e_i)^T\mathbf{Q}^\dagger(\mathbf{p}-e_i)
\\
&= -\frac{1}{2}(\mathbf{p}-e_i)^T\Omega(\mathbf{p}-e_i)
\\
&= -\frac{1}{2}\left(\mathbf{p}^T\Omega\mathbf{p} - 2e_i^T\Omega\mathbf{p}\right)
\\
&= \frac{1}{2}\mathbf{p}^T\Omega\mathbf{p}=\var_{\omega}(\mathbf{p})
\end{align*}
This confirms that (i)$\Leftrightarrow$(iii)
\\
\textbf{(i)$\Leftrightarrow$(iv)} Using the definition of the effective resistance matrix, we can write
\begin{align*}
&\Omega\mathbf{p} - 2\var_{\omega}(\mathbf{p})\mathbf{u}
\\
&=\left[(I-\mathbf{u}\mathbf{u}^T/n)\zeta - 2 Q^\dagger \mathbf{p}\right] 
\\
&\hphantom{=}- \left[2\var_{\omega}(\mathbf{p})-\zeta^T\mathbf{p}-\mathbf{u}^T\zeta/n\right]\mathbf{u}.
\end{align*}
Since the first term in square brackets is in $\mathbf{u}^{\perp}$ while the second term is parallel to $\mathbf{u}$ we know that (i) holds if and only if both terms are zero, and thus
$$
2Q^\dagger\mathbf{p}=(I-\mathbf{u}\mathbf{u}^T/n)\zeta \text{~and~}\var_{\omega}(\mathbf{p})=\zeta^T\mathbf{p}-\mathbf{u}^T\zeta/n
$$
which is satisfied if and only (iv) holds, i.e. when $\mathbf{p}=\tfrac{1}{2}Q\zeta+\mathbf{u}/n$ and with the corresponding variance $\var_{\omega}(\mathbf{p}) = \tfrac{1}{4}\zeta^TQ\zeta + \mathbf{u}^T\zeta/n$.
\\
\textbf{(i)$\Leftrightarrow$(v)} Using Fiedler's identity \eqref{eq: Fiedlers identity} to encode (i), we can write
$$
\begin{pmatrix}-2\var_{\omega}(\mathbf{p})&\mathbf{p}^T\\\mathbf{p}&-\tfrac{1}{2}Q\end{pmatrix}\begin{pmatrix}
0&\mathbf{u}^T\\
\mathbf{u}&\Omega
\end{pmatrix}=I\Rightarrow \mathbf{p}\mathbf{u}^T=\frac{1}{2}Q\Omega -I.
$$
Hence, taking the $i^\text{th}$ diagonal of this expression and using the fact that $\omega_{ii}=0$, we find
$$
p_i = 1-\frac{1}{2}\sum_{j\sim i}c_{ij}\omega_{ij}
$$
which proves (i)$\Leftrightarrow$(iv) and thus completes the proof.\hfill$\square$
\\
We now derive Proposition \ref{propos: relation between maxvar distributions} which relates the solution $\mathbf{p}$ of \eqref{eq: 2 local optimality condition} for the full node set, to the solution for subsets.
\\
\textbf{Proof of Proposition \ref{propos: relation between maxvar distributions}:} We let $\mathcal{V}=\mathcal{N}\backslash\lbrace x\rbrace$ and from Fiedler's identity for the resistance matrix $\Omega_{\mathcal{V}\mathcal{V}}$ we find
$$
p'_i = e_1^T\begin{pmatrix}
0&\mathbf{u}^T\\\mathbf{u}&\Omega_{\mathcal{V}\mathcal{V}}
\end{pmatrix}^{-1}e_i.
$$
where $\mathbf{p}'$ solves condition \eqref{eq: 2 local optimality condition} for the set $\mathcal{V}$. This inverse resistance matrix is the inverse of a submatrix of the matrix $\left(\begin{smallmatrix}0&\mathbf{u}^T\\\mathbf{u}&\Omega\end{smallmatrix}\right)$ which appears in Fiedler's identity \eqref{eq: Fiedlers identity} for the full node set. Using the Schur complement formula for block-matrix inversion (see also \eqref{eq: Schur complement}), we find that
$$
\begin{pmatrix}
0&\mathbf{u}^T\\\mathbf{u}&\Omega_{\mathcal{V}\mathcal{V}}
\end{pmatrix}^{-1} = \begin{pmatrix}
-2\sigma^2&\mathbf{p}_{\mathcal{V}}^T\\\mathbf{p}_{\mathcal{V}}&\frac{-1}{2}Q_{\mathcal{V}\mathcal{V}}\end{pmatrix}
+ \frac{2}{k_x}\begin{pmatrix}p_x\\\tfrac{1}{2}\mathbf{c}_x
\end{pmatrix}\begin{pmatrix}
p_x\\\tfrac{1}{2}\mathbf{c}_x
\end{pmatrix},
$$
with vector $(\mathbf{c}_x)_i=c_{ix}$. Consequently, for $p'_i$ and $\sigma'^2$ we find
$$
\begin{dcases}
p'_i = p_i + c_{ix}/k_x\\
\var_{\omega}(\mathbf{p}') = \var_{\omega}(\mathbf{p})^2-\frac{p_x^2}{k_x}
\end{dcases}
$$
as required.\hfill$\square$\\
\textit{Remark:} To illustrate the convenience of Proposition \ref{propos: relation between maxvar distributions} we note that it allows for a particularly simple proof of Lemma \ref{lem: p for weighted star}. In any weighted star on $n$ leaf nodes, the solution to \eqref{eq: 2 local optimality condition} gives $p'_i=1/2$ for each leaf node and $p'_0=(2-v)/2$ (e.g. by Proposition \ref{propos: equivalent characterizations} (v)). Thus if we take the Schur complement with respect to $\lbrace 0\rbrace^c$ we find by Proposition \ref{propos: relation between maxvar distributions} that $p_i=1/2-(v-2)k_i/(4m_\mathcal{V})$ as required.
\end{document}